# Smart Parking: IoT and Blockchain

Abdul Wahab

Dissertation 2019

DEPEND Erasmus Mundus Joint MSc in Advanced Systems Dependability

Department of Computer Science,

Maynooth University,

Co. Kildare, Ireland.

A dissertation submitted in partial fulfilment
of the requirements for the
Erasmus Mundus MSc Dependable Software Systems

Head of Department: Dr Joesph Timoney

Supervisor: Dr Phil Maguire

June 20, 2019

## Declaration

I hereby certify that this material, which I now submit for assessment in the program of study as part of Erasmus Mundus MSc Dependable Software Systems qualification, is entirely my own work and has not been taken from the work of others - save and to the extent that such work has been cited and acknowledged within the text of my work.

**Abdul Wahab**

**June 2019**

**Acknowledgements**

I would like to thank my supervisor Dr. Phil Maguire for his guidance, patience, feedback and support during the project. His confidence and interest in this cutting-edge field and the project was always a source of motivation for me. I would also like to thank Dr. Jodeph Timoney for his diligence and dedication to the Erasmus Mundus MSc in Dependable Software Systems programme. Lastly, a sincere thank you to those who volunteered in research, provided guidance on my software development strategy and gave their valuable feedback.

## Abstract

Distributed ledger technology and IoT has revolutionized the world by finding its application in all the domains. It promises to transform the digital infrastructure which powers extensive evolutions and impacts a lot of areas. Vehicle parking is a major problem in major cities around the world in both developed and developing countries. The common problems are unavailability or shortage of parking spaces, no information about tariff and no mean of searching availability of parking space online. The struggle doesn't end even if an individual finds a spot, he is required to pay in cash. This traditional and manual process takes a lot of time and causes a lot of hassle. In this paper, we provide a novel solution to the parking problem using IoT and distributed ledger technology. This system is based on pervasive computing and provides auto check-in and check-out. The user can control the system and their profile using the app on their smartphone. The major advantage of the system is the easy and online payment method. User can pay for their parking tickets using their credit card from their smartphone app. This decreases their hassle of carrying cash and coins for purchasing parking tickets. Smart Parking will optimized the parking mechanism, save time, reduce traffic and pollution, and provide an enhanced user experience. It is robust, secure, scalable and automated using the combination of cutting-edge technologies.

# Table of Contents

# List of Figures

# List of Tables

# Chapter 1

# Introduction

Smart parking has always been a strategic and crucial issue to implement due to its economic benefits. Despite the backing of extensive research and implementation models, it has not yet been implemented even in the metropolitan cities of developed countries. Parking realigns transportation and roads utilization in the cities. It acts as an asset and serves as a source of revenue. To realize, how much land parking utilizes in the metropolis cities around the world, we look the Table 1 below.

| **Cities** | **Parking Coverage (in terms of land)** |
|---|---|
| Los Angeles | 81% |
| Melbourne | 76% |
| New York | 18% |
| London | 16% |
| Tokyo | 7% |

*Table 1 Parking land utilization (Lin , Rivano, & Mouël, 2017)*

This table shows that parking makes up a significant area of a city and by improving the legacy parking mechanism using IoT and blockchain can significantly affect the time and effects of the driver. Many cities have begun working on the concept of smart parking which is to search for parking and automate the process of parking ticket using information technology. (Lin , Rivano, & Mouël, 2017)

The parking problem is one of the major and crucial problems. In the busiest hours, it becomes very difficult to find a parking spot. In case, you find one, securing the place is another hassle. Almost all the parking terminals accept coins and provide hourly based parking. The end user needs to purchase a parking ticket and place it on the vehicle's windscreen. In case the user needs more time, he will need to come back and purchase another parking ticket and replace from the one already placed. This is common, and we cannot properly estimate the amount of time we will take at a place. The current parking issues requires a smart parking implementation using modern technology.

In the current era, technology is transforming our world exponentially. The major focus has been towards automation, security and smart decision-making devices. Ubiquitous computing is a modern computing concept that is based on a pervasive paradigm. IoT devices used to implement ubiquitous computing, have gain wide popularity for being wearable, powerful and highly usable. It has been actively adopted both in person and on the enterprise level.

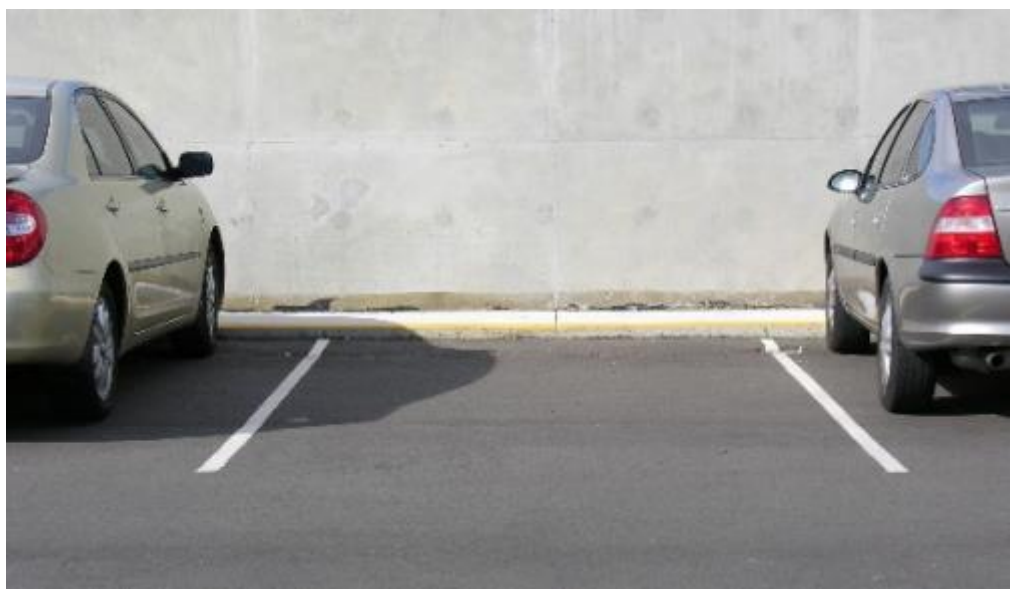

*Figure 1 Parking Area*

IoT is a futuristic technology that can form a network of one or more devices over the internet. It has multiple sensors, a computing device, other hardware, and a programmable disk. The sensors and hardware are used to fetch information and pass it to the computing device to decide. It enables a machine to machine interaction and is usually used to spread out the information in real time to enable decision making, efficient control, and automation. It has found its application in

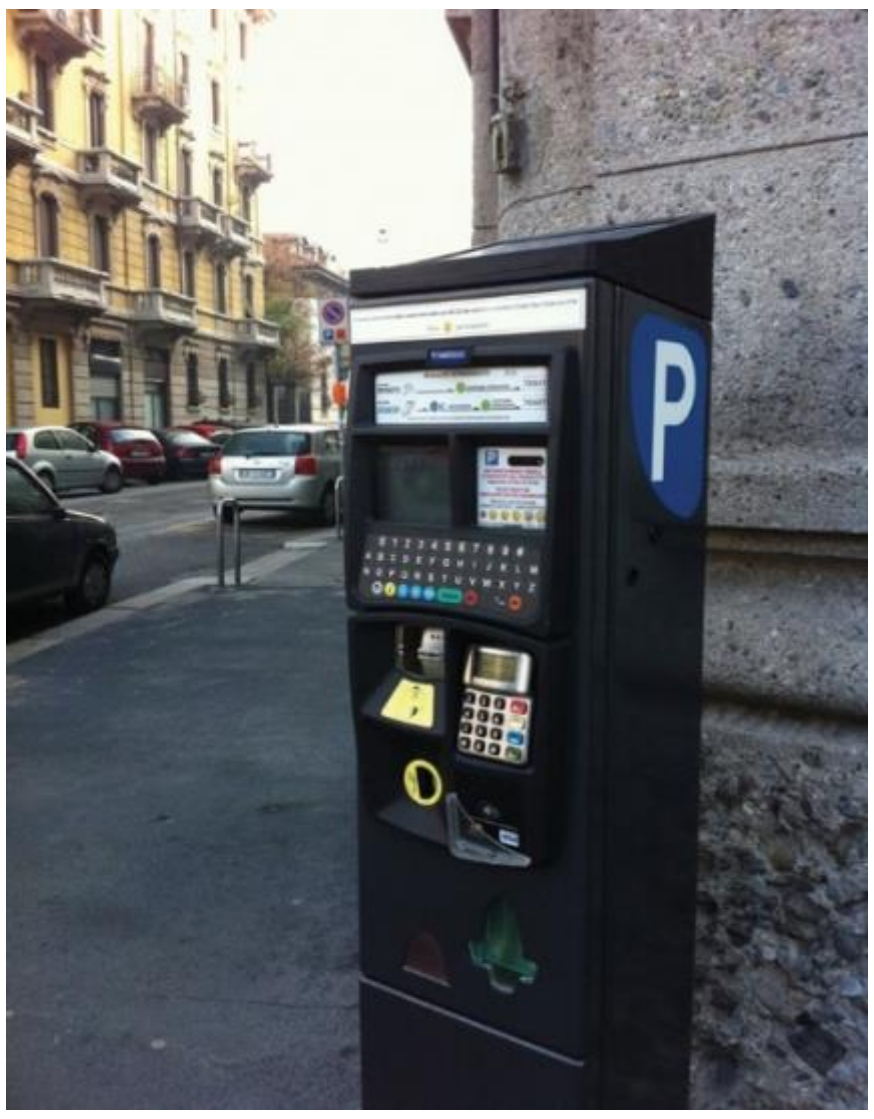

*Figure 2 Parking Terminal (ITS International, 2012)*

everything including coffee maker, cell phone, refrigerator, wearable devices, and other home appliances. Not only that, IoT is being adopted by the industry as well, hence the term Industrial IoT. It is being used in factories and workshops to automate their process and streamline their workload. It has also found its major application in fault detection. As per the technology analysts, IoT is rapidly integrating without technology and is expected to connect 26 billion devices to the internet by 2020. Due to the endless possibilities, it has coined several new technology paradigms such as people-to-things, things-to-things, and people-to-people. (sticam13, 2018)

Blockchain has transformed and disrupted the traditional business model to achieve significant business benefits. It offers transparency, decentralization, and security. It is a type of distributed ledger where all the participants keep the copy of the ledger which is only updated when all agrees to it, through a consensus. This guarantee data accuracy and form data trust. All the data is stored in the form of an encrypted transaction. It forms the basis of security which is crucial for financial, healthcare and government institutions. It is finding its use case in all the discipline of the industry especially asset tokenization. It has also been implemented in the energy market to record metering of electricity generated through solar panels. In Switzerland, the passport and identity have been replaced by digital identity on the blockchain. The reason for the rapid adoption of this technology is its internal working that guarantees security and immutability of data. (Hooper, 2018) (Su, 2018)

IoT sensor can be used to setup auto check-ins and check-outs once the vehicle is in the parking area and has been parked. This will not only speed up the check-in and check-out process but will save a lot of hassle faced by the end user. When people can spot parking quickly, it will reduce traffic congestion. The parking logs, check-ins, and check-outs will be stored on the blockchain for data immutability and security. As these logs will be the basis for estimating the duration and costing of the parking ticket. We need a system that cannot be compromised and is the foundation of trust and data immutability. Additionally, blockchain is decentralized that makes the system available even if one or more node breaks down.

## 1.1 Motivation

This thesis focuses on and attempts to streamline the payment process for the parking ticket. Currently, all the parking tickets in all the major cities require you to pay in coins, and at a few places, we are required to pay the exact amount. This is a cumbersome process and people don't have exact coins all the time. Another issue with this process is the proper utilization of parking ticket. Current tickets are for at least an hour where most of the time all you need is a parking spot for just a few minutes. This causes people to pay more for less.

The most inefficient and time-consuming process is to find a parking spot. All we do it to roam around different areas and squares in the city to check if the parking is available. We need a system that could tell us where the nearest available parking spot is. This system is also very inefficient as we do not know the available parking spaces and remain unused in times of high peak hours. If a system is developed that tell all the available parking spaces, it will not only allow efficient use of parking spaces, more parking tariff generation, and more people get the parking easily and effectively. We need a system using modern technologies to ensure all the available parking spots are used up effectively. This is will not only reduce the parking tariff as all the parking spots will be used up effectively but also save the time of people as they will exactly know where to park their vehicle. (Ibrahim, 2017)

With the current manual parking system, the vehicle remains on ignition as it finds a parking spot. The driver searches at multiple places to secure a spot. Then next are parking terminals, the vehicle on ignition must wait until the driver pays a certain amount and take the ticket. Same waiting occurs at the checkout process where the driver submits the ticket back to the terminal. Again, the time of check out depends on the driver and is usually prolonged as the driver is sometimes unable to find the parking ticket. During this whole process, the vehicle remains on the ignition and leads to the formation of emission gas. This is an environmental issue and can be minimized by automatic ticket allocation for parking. There is a need for a system where check-ins and check-outs are automated to reduce waiting time at a parking terminal. The reduce waiting time will also contribute to reduced fuel consumption. A system that can search for the parking space will also give clear direction to the driver about the exact location. This will reduce traffic congestion as the driver will not cruise around the city for parking space. (Inanc & Pala, 2007)

Another issue with the current traditional parking system is the fraud committed by people. Many people take chances and use the perused parking ticket. They place the same ticket daily on their windscreen in the hope that the office will not check it. Now it is not possible for the city administration to check all the parking tickets at every spot. So, it is verified randomly, and a lot of people run away with fraud. This cost a lot to the city council and the manual inspection of the ticket is also very costly. There is a need for an automated system that can avoid such fraud. An automated system will also reduce the need and cost of manual inspection of the parking ticket.

## 1.2 Project Objective

In this thesis, we shall provide a solution for the parking problem in the metropolitan city. The major objective of the project is to devise an automated process using pervasive computing. There should minimum intervention and input from the user and the whole process should be self-automated. This is still an untapped area and a lot of work has been proposed to automate the process. This thesis will review all the existing development in this area to date and provide its own solution based on cutting edge technologies.

This thesis also focuses on and attempts to streamline the payment process for the parking ticket. The project has implemented a process of auto check-in and check-out for the parking time log. This will help us determine the total duration of the time, the vehicle was parked. The precise amount of duration will allow us to precisely charge for the parking spot. The thesis will also work on different payment options available to pay for the parking tickets.

Another objective of the thesis on a broader perspective is to provide a framework for solving daily life problem using cutting edge technologies. IoT and blockchain have huge potential for digital transformation and enterprise integration. These technologies can be used to solve a variety of use cases and provide efficient industry solutions. The thesis attempts to solve the parking problem using these set technologies by exploring how they can integrate with each other for leveraging better value. In this paper (Sisinni, Saifullah, Han, & Jennehag, 2018), the author has done a similar job. On a broader perspective, they tried to integrate IoT with industrial application, Industrial IoT. They tried to integrate ubiquitous to enable smart decision and to create pervasively connected infrastructure. They tried to pick up a use case for industry and tried to solve it using pervasive computing paradigm. Similarly, in this these, we have picked the problem of parking issue and tried to solve it using IoT and blockchain. On a broader perspective, we are trying to see how IoT and blockchain can work together to solve enterprise problems and analyse their benefits and challenges.

This thesis will also explore how can we integrate ubiquitous computing and explore how its merit and demerits. Ubiquitous computing is a breakthrough in the field of computer science and offers efficiency, effectiveness, and optimizations in computing processes. Additionally, it will also explore design decisions regarding multiple technologies to derive the reasoning for choosing one for our use case. We will also explore the usefulness of these technologies.

## 1.3 Dissertation Structure

The remaining chapters are structured as follows: In chapter 2, we will provide a literature review of the existing solutions and use cases on the said problem. This section will also highlight the most relevant solution and contribution in this field up to date. We have named multiple contributions under similar heading to understand the progress on the contribution over time by multiple authors under the same domain. Chapter 3 will introduce the major technologies and provide background on it. It will also provide concepts and use cases about combining these technologies for digital transformation. This section also compares some of the powerful technologies and strategies, and reasons why we selected one over another. Furthermore, it will discuss the reasoning behind the design decisions that lead to the selection of a set of technologies and architectural decisions. The section also discusses the architecture of the project and how different components connect with each other. The design decision will show the usefulness of multiple technologies and how are they suitable for our use case. Finally, this section will discuss the solution and how the whole system is operated. Chapter 4 finalizes the thesis and outlines the conclusion of the project. It will also review the future work and further opportunities in this domain. We will also discuss the limitation and issues with the current project that can be rectified in the future.

# Chapter 2

# Related Work and Background

Vehicle parking is the biggest problem of cities nowadays, due to an increasing amount of vehicles day by day. To enhance the parking resources and to decrease the efforts of a driver to search for a parking spot, many new technologies are implemented in the past few decades and IoT is one of them. It connects things together through sensors and data collection is done and communicated via the internet (Khanna, 2016). Government has implemented Parking Guidance and Information (PGI) system at major roads and streets that tell drivers about vacant spaces. Variable Message Signboards (VMS) are used to display vacant parking spot information (Renuka, 2015). To find a parking spot is not the only issue. If a driver gets to succeed in finding a spot, he must pay for that parking. There is a ticketing process driver must follow to have legal parking otherwise he will be fined by police as per law.

This Vehicle parking problem was addressed by many researchers as it is a major problem of big cities and its cost is very high in terms of fuel and time consumption. These solutions can be categorized as parking reservation models and pricing models where the major concern is parking cost (Kotb, Shen, Zhu, & Huang, 2016).

## 1.1 Parking Reservation Models

To solve this major problem, many solutions were provided and implemented in different cities and countries around the world. Some cities have implemented IR sensors at each parking spot that sense light wavelength of its surrounding or measure heat emitting from nearby objects (Renuka, 2015). The sensor is red if the spot is occupied and turns to green when it's vacant. In (S. Mendiratta, 2017), they have advanced this technology by connecting those sensors with an IoT device to collect the data and this data is transmitted through Wi-Fi to the driver or parking management.

In (Renuka, 2015), the authors proposed an android based smart parking system with real-time allocation method that reduces the stress of driver and reduces air pollution and global warming. They implemented IoT to get the vacant parking slot information and RFID tags for automatic billing process. In their model, they have extended the IR sensor system and combine it with an android application to work smartly. The system consists of three steps. First, driver request for the parking space using his android application, driver's details, car license number, etc. are sent to the Driver Request Processing Centre (DRPC). Smart Parking Allocation Centre (SPAC) collects all drivers' requests and based on two constraints; the distance between driver's current location and vacant parking slot and the cost of parking, a slot location along with distance and cost is sent back to the driver. If driver agreed to the location, he accepts that location and the allocated parking slot is updated from vacant to reserve in Parking Resource Management Centre (PRMC) and the IR sensor LED light change from green to red. If the driver is not satisfied with the allocated location, the request is again sent to the system. When a car enters the parking spot, timer for the duration of car parking gets started and the moment the car leaves the spot it stops and based on duration, the cost is displayed. The cost is paid through RFID tags using RFID application.

In another study (T. N. Pham, 2015), the author proposed a cloud-based smart parking system using the concept of IoT. In this system, they have developed a network of all parking areas, and data is transferred to the data centre which is on the cloud. This data contains a GPS location of the vehicle, the distance between the vehicle parking area within the network and the free spot in each parking area. They have used RFID tags for the authentication of user information and to check the vacant parking spots in each car parking area. The user request for the parking spot and the available spot is notified to him via android application based on his current location and the location of car park sorted by cost. This notification includes the address of the car parking and the direction to it. Once a parking spot is allocated to a user, its status is updated to “pending” in the cloud-based data centre. If a user does not park his car for a certain period, its status is updated to “available” again. When the driver reaches the parking spot, its credentials are verified using an RFID tag or by scanning user card. If the user is authenticated, he can park his car on a specified spot. This model is an advanced model in comparison to previous ones because it considers load balancing to the parking areas as it creates a network of parking areas and all the data is on cloud, so it can easily redirect a driver to another nearby parking area with minimum cost if a parking spot is not available at the current parking area.

In (Z. Mahmood, 2019), the authors have proposed face and vehicle detection algorithms. The model suggests implementing a CCTV camera at the entry and exit gate of the parking lot. When a car enters it will detect the face of the driver and other vehicle information such as car number and saves it in a database. AdaBoost along with Haar features is used to detect and locate vehicle and face of the driver. They have used three baseline face recognition algorithms (local binary pattern (LBP) based face recognition algorithm, AdaBoost-linear discriminant analysis (LDA) face recognition algorithm, and principal component analysis (PCA) based face recognition algorithm). When a car leaves the parking lot, at the exit gate it will again capture the images and compare them with those taken at the entry gate to verify the identity of the driver. This model can help in calculating the time duration of parking and is thief prove as it checks driver’s identity but the main problem with this model is to take a correct image of driver as well as vehicle because taking an image of moving obstacle is a big challenge. Secondly, it is not necessary that the sitting position of the driver at the entry point is the same as that on the exit point. This will also make a comparison of images difficult task. Different experiments prove this model is effective and reliable under different lighting conditions.

In the paper (Kotb, Shen, Zhu, & Huang, 2016), they have proposed a model (iParker) to the parking problem by offering parking reservation with the lowest cost. Their model is based on mathematical modelling using mixed integer linear programming (MILP). Due to reserved parking, time and fuel cost of a driver can be saved and high utilization of parking slot can be done due to the known availability of parking spot with results in higher revenue for parking management. It has a real-time dynamic pricing model where pricing increase or decrease is proportional to the utilization of parking space.

In major cities, a large parking area is owned by the private sector which is vacant most of the time. In (L. Zhu, 2018), authors have proposed the idea to use those private parking spots along with public parking. Due to low utilization rate and high maintenance cost, owner of these private parking areas is willing to provide their parking spots to the public on low charges and with the increase in the use of smartphones it is easily possible. The major problem with this idea is to privacy issue of both driver and supplier. In (L. Zhu, 2018), they have considered all these issues and proposed the model accordingly. The model consists of four entities; trusted authority (TA) whose responsibility is to register supplier and driver and distribute keys, a server who get the report from both driver and supplier and sends the best matching result to the driver and manage the payment process, a driver and a supplier. To secure the identity of both driver and supplier, the authors have used short randomized group signature method. It is a cryptographic technique to create a signature using a bilinear pairing and digital signature. This signature key is used as an identity for both. To get faster results, all the parking locations are saved on HashMap in the cloud. A binary tree is used for HashMap so when a driver query about the parking lot, it takes O ($\log_2$ Num) time to send back matching result sorted by price. To secure the location of the driver they have used cloaking technique. To pay for the parking fee, E-Cash mechanism is used. Again, the identity of both driver and supplier are hidden from each other in the payment process. The driver sends the coupon to supplier generated by blind user signature keeping driver's identity secure and is encrypted using the supplier's public key, so no other supplier can decrypt it. The supplier sends this coupon to the server to check its validity. The server verifies coupon validity using the supplier's public key so the real identity of both driver and supplier are protected.

## 1.2 Pricing Models

In many cities, the government has installed a ticketing system in the parking area. There are different payment processes, in some cities driver gets a parking ticket at the entrance gate and when they exit from the parking lot, their ticket gets scanned and the calculated amount has to be paid (Z. Mahmood, 2019). In other cities, the driver parks his car, go to the paying station, predict the duration of parking and pay in coins to the machine accordingly. A ticket is generated by a machine which must be placed on the dashboard of the car. A patrolling officer checks each car and if any car does not have a ticket or has exceeded the time mentioned on the ticket will be fined as per law. In this way, the car owner is restricted to time so that any new driver can find parking easily. But this approach is not much efficient, time-consuming and has many drawbacks.

- The required amount of coins is necessary because machine take fix amount and credit card is not supported in most spots.
- Driver must predict for his parking duration in advance. If a driver predicts more time than actual parking time duration, the money for over predicted time is lost.
- If a driver predicts less time, then he must come and perform the whole ticket taking process again which is again a time-consuming task otherwise he will be fined by the officer. (Pérez-Martínez, 2013)
- If a driver misses that ticket, he must take the ticket again.
- If a driver has paid for parking and he must move to some other parking area, he must pay for that parking spot again although the previously issued ticket is not expired.

To solve this problem, pay by phone applications are implemented in many cities to save the time of their citizens. In this, the user creates an account and can pay electronically. When a user parks his car, he enters the license number, car parking location, and the expected time duration of parking. After entering these details, the transaction is made according to the payment time selected by the driver while creating an account. The driver can change the time duration any time and the money that was charged against unused time is refunded back to his account or if he increases the time duration another transaction corresponds to the increased time is made from his account. Patrolling officers can check the paid/unpaid status of the car by another mobile application where they enter car license number and status is shown to them. But this type of application has privacy issues as all the driver details can be viewed by a patrolling officer.

In (Pérez-Martínez, 2013), authors have proposed a very general model that hides the identity of the driver using E-Cash. In the model, the user first gets the electronic cash from the bank by using any existing protocol. To pay for the parking lot, the user sends an activation message to the RFID tag located in the vehicle. The tag uses a hash function to generate a random number and that number is sent back to the driver which is used to pay for the parking spot. The E-Cash paid by the driver is verified by the bank and the original money is deposited to the service provider. The money transferred to the service provider is converted in the form of time duration for parking. This time duration is saved to the database and the expiry time is notified to the user. The model provides anonymity, remote payment, and intractability.

In this paper (R. Garra, 2017), the authors focused on the payment method of the car parking and introduced electronic coin (e-coin) for the payment. In this proposal, the driver needs to install an application on his mobile phone and an RFID and NFC enabled device on his car. This device is cryptographically connected to mobile app sharing some secret key. The electronic coin is stored in an electronic wallet and is top up when finished. When a driver parks his car, he logs in to the app and initiates the parking process. The initial charges are charged and paid using e-coins to the system server. The payment is made for a very short time span (e.g. 10 minutes) after each 10 minutes the app contacts the system server and pays for the next 10 mins electronically. This process continues until the driver removes his car from the parking spot. All the payments are saved in the database in the form of a timestamp. If a driver finds any miscalculation or unfair transaction, he can complain through the app and after processing his complain, If the complaint is correct, e-coin will be deposited to his wallet again i.e. the system is saved against double spending and e-coin forgery. If the electronic wallet gets empty, the user can recharge it by buying new e-coins using his credit card or through any other service such as PayPal. For the parking authority, the checking of the parking status of each car is also made easy. The patrolling officer comes near to the car and scans the RFID tag by a mobile device carried by him, the request is sent to the server and only the status of payment of parking i.e. if the driver has paid for the parking or not is sent to the parking officer. Due to privacy issues, no personal or car information is sent to the officer.

# Chapter 3

# Technologies and Decisions

## 1.1 Technologies used

The goal of the project is to automate the parking process and find available parking spots online. The project is based on the use of IoT device for auto check in/check out from the parking lot. It uses blockchain to securely save all the vehicle parking logs and generates monthly bills. The end user uses a mobile app to register on the system, set up its vehicle, view parking logs and process bill payment. This whole system requires a lot of technologies and infrastructure to make it functional. This section discusses all the major technologies used.

### 1.1.1 Docker

Docker (Docker , 2018) (Anderson, 2015) is an opensource containerization technology that enables the users to create, build and deploy applications as Docker containers. Since its inception in 2013, containerization transformed the enterprise infrastructure and started becoming an essential part of it. The technology eliminates the dependencies between application and the underlying physical infrastructure which provides portability and efficiency. It is a successor to virtual machines. Unlike virtual machines, it doesn't create a virtual operating system for the underlying application or software to execute. It uses the host OS kernel to run the application and all its dependencies. This significantly increases performance and reduces the size of the application.

From a developer's perspective, docker allows the developers to package up their applications, all its components, and their dependencies and ship it out as one package. This package is called a docker image. That image can easily be distributed and deployed on a server or another machine. Docker is fast and unlike virtual machines, it boots up quickly. Docker is multi-platform and that makes your container image launchable on any operating system. (Zhang, et al., 2018)

*Figure 3 Docker Logo (Docker , 2018)*

To setup docker on your machine, you need to install its binaries from its official website. The docker software allows you to create, build, deploy and share your container images. Docker Compose (Docker, 2018) is another tool that allows you to define and run multi-container docker applications on a single machine. It also enables the user to create a network among the containers so that they can exchange data and communicate with one another. It is defined by YAML file which contains the container images, network information, and other required properties. A basic YAML file looks like the one in the figures. In this file, you can define multiple containers, their dependencies, and their order of spawning. You can also extend multiple files for reusing their container definitions.

Docker Swarm (Docker, 2018) allows a group of machines running Docker containers to join a cluster and make a swarm. In a swarm, you can execute commands all the machines in a cluster using a swarm manager. All the other computers are treated as nodes. This technology enables us to set up an environment and architecture for distributed computing. One common use case for such a setting is Hyperledger Fabric blockchain. The nodes in the blockchain have a complex set of software so they are packages as docker images for flexibility and ease of use. All these nodes are clustered by docker swarm to communicate with each other and form a distributed network.

### 1.1.1 Hyperledger Fabric

Hyperledger Fabric (IBM, 2016) is a permissioned blockchain implementation infrastructure developed by IBM and hosted by The Linux Foundation. It provides an application development framework on the top of a modular architecture. It supports plug and plays model of components where consensus and membership service are modular. It also supports application logic development using a smart contract which is available in Go language. It is a scalable blockchain architecture for developing and executing distributed applications.

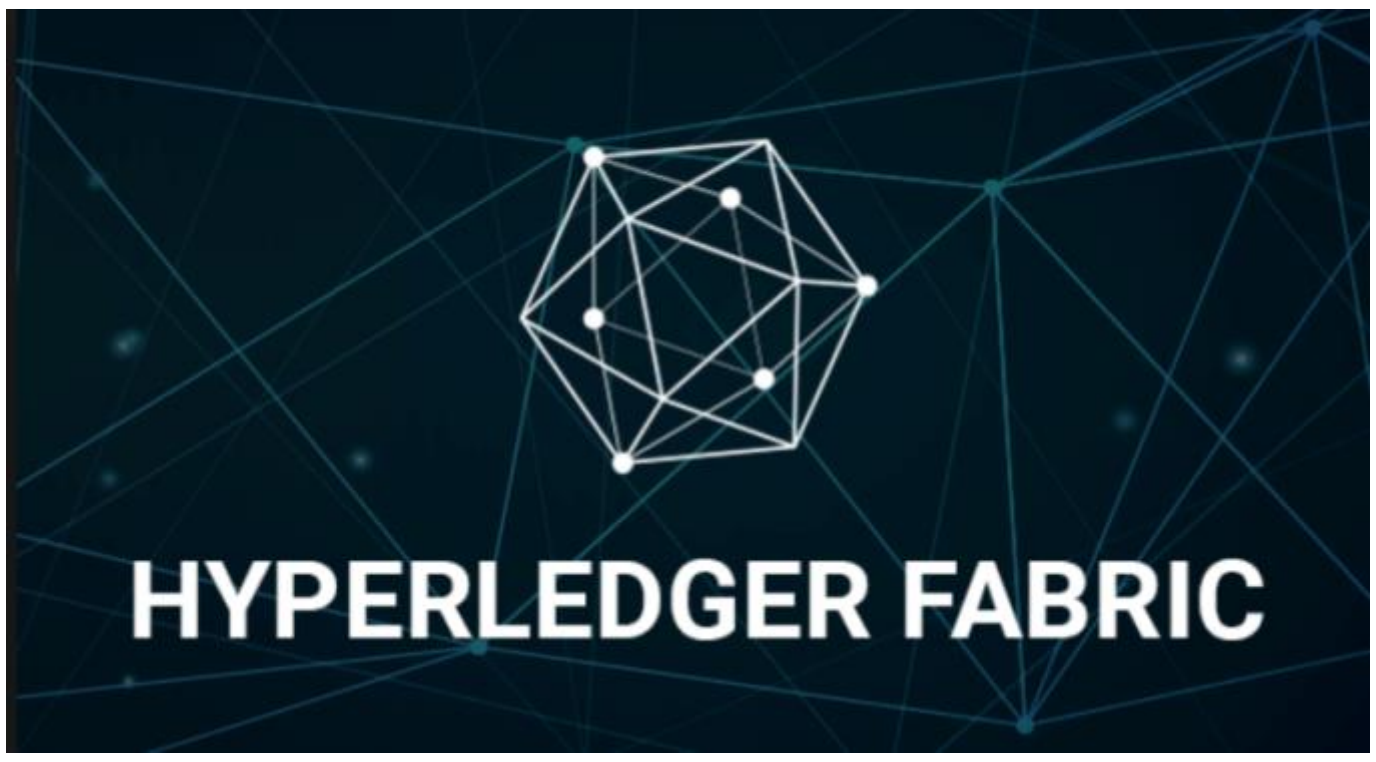

*Figure 4 Hyperledger Fabric Logo (Hyperledger, 2018)*

A fabric blockchain network is comprised of two types of nodes; Peer node and Orderer node. Peer node is responsible for executing chain-code, endorse transactions, manage ledger data and interact with end-user application. An Orderer node maintains the consortium and consistency of the ledger. It communicates back and forth with the peers and provides endorsed transactions to commit to their ledger. The Fabric utilizes a trusted Membership Service Provider (MSP) to leverage permission-model architecture with standard and secure identity management. (Hyperledger, 2017) (Cachin, 2016)

Data security and transparency is a big challenge for any distributed system. Fabric incorporates channel implementation to achieve this feature. A channel is a permissioned and secure data path. Only the peer who has joined the channel can access and operate on the data. This is useful for a variety of use cases where data security among peers is an essential requirement. If a group of peer form and join a channel, the ledger data of that channel is only accessible to them. (Cachin, 2016)

The Fabric also secures identity management for your project. The project is shipped with the pluggable certificate authority to generate certificate and key for your application participants. This manages their access and actions in the blockchain network. You can also plug your custom certificate authority service which generates certificates based on ECDSA. (Cachin, 2016)

A Distributed ledger, which is the heart of the blockchain network, is managed by a key-value pair database. It is distributed in nature and is replicated across all the nodes of the network. Smart contracts are used to create data on the ledger. It serves a mean to automate transaction execution on the network. Then it is the responsibility of the consensus to maintain and synchronized the data across the network through a collaborative process. (Hyperledger, 2017)

### 1.1.2 Hyperledger Composer

Hyperledger Composer (Hyperledger, 2017) (Dhillon, Metcalf, & Hooper, 2017) is a toolkit for simpler and smooth development of smart contracts and distributed applications on the top of Hyperledger Fabric. It aims at accelerating blockchain application development to solve business problems using distributed ledger technology.

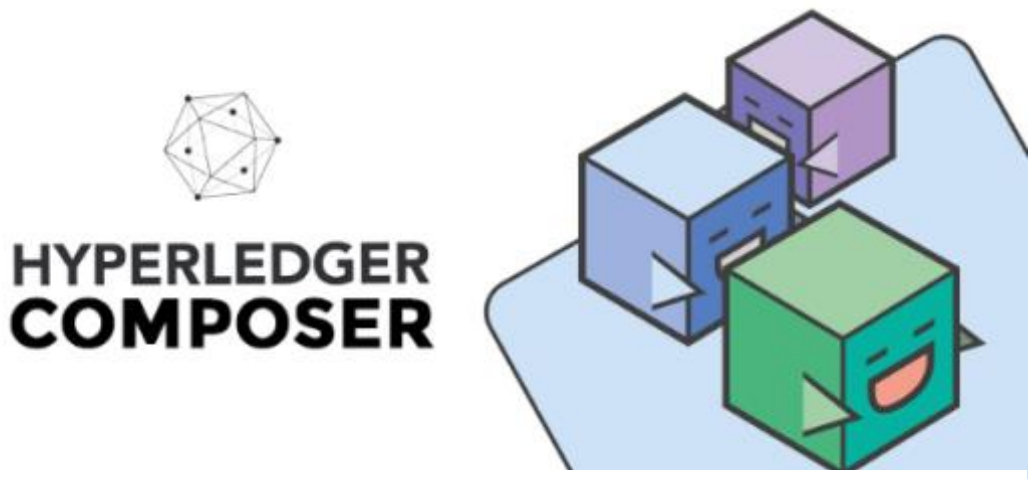

*Figure 5 Hyperledger Composer (Djaja, 2018)*

Composer utilized fabric blockchain architecture and runtime to test and deploy an application. Applications developed using composer and are executed on fabric runtime if you are running it locally on your machine. The Composer also provides an online playground where you can develop and test your applications without any fabric runtime dependency. (Hyperledger, 2017)

A composer application consists of assets, participants, transactions, and events. An asset could be any tangible or intangible good, property or service. Participants are users which could both administrative users such as network admin or end users. Transaction is conducted on the assets by participants. They interact with the assets to fulfil business logic. Event are created and dispatched on transactions which can be listened by any third-party application. (Hyperledger Community, 2019)

To control access to assets among different participants, the composer utilizes access control language (ACL) to implement secure data architecture. It gives an administrative control what asset and transaction a participant is authorized to access and execute in a business network. (Hyperledger Community, 2019)

The Query language is another important component of the composer that allows the user to query data from the ledger. This lets you compose SQL statements that you can execute on the ledger. The results can be sorted and filtered using composer query keywords. The query language can be used to fetch and modify results from the ledger. (Hyperledger Community, 2019)

Composer CLI provides commands to deploy, update and start a Business Network Application (BNA) on a network. It can also expose the business network as RESTful API on Open API standards. This APIs can be used by other applications to communicate with the network application.

### 1.1.3 Hyperledger Explorer

Hyperledger explorer (Hyperledger Community, 2019) is one of the Hyperledger project aimed at accessing, invoking and querying transaction, chain code and block data on the blockchain network.

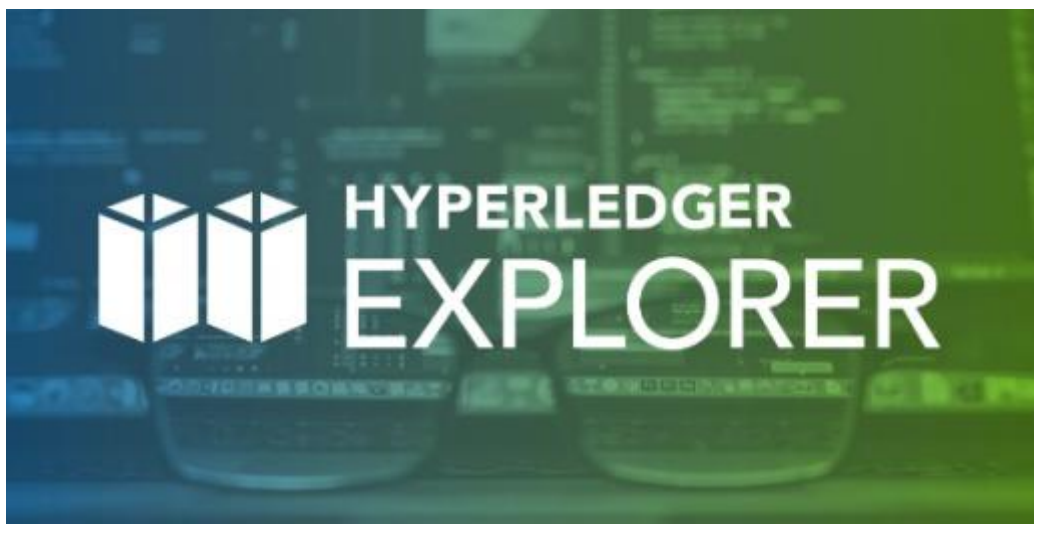

*Figure 6 Hyperledger Explorer Logo (Hyperledger, 2019)*

It is powerful, easy to use and opensource web app that is connected with the blockchain network to displays the relevant information stored in its ledger and watch activities on the blockchain network. It also demonstrates the list of connected nodes, their states, and network information. (Longstaff, 2019)

To install Hyperledger explorer you will need to install all of its prerequisite that includes Node.js and PostgreSQL. It comes with configurable JSON files to connect with your local PostgreSQL database and fabric blockchain network. It monitors the network in real-time which means you see the block creation on the web as you perform transactions on the ledger. It has a backend that connects with blockchain service using web sockets. It allows Hyperledger Explorer to connect with multiple components of the blockchain network. The PostgreSQL is used to store all the necessary information about block creation and transition details to querying purposes. It also manages a security repository to ensure authorized access to the Explorer web interface. (Supinnapong, 2018)

In addition to block search and monitoring service, Explorer allows you to visualize and understand the raw blockchain data which is difficult to read. It does it by integrating charts, graphs, and multiple analysis tables into its web dashboard.

### 1.1.4 IBM Loopback 4

Loopback (IBM, 2019) is an extensible and opensource Node.js platform for developing web APIs. It a modern, mature and enterprise-ready JavaScript framework for crafting APIs using Model-View-Controller (MVC) standard. It offers the rapid creation of end-to-end APIs using scaffolding which makes developing and testing application faster and simpler. (Create REST APIs in minutes with LoopBack 4, 2018)

To start using Loopback, the user first needs to install CLI tools. Loopback CLI create a template for project and boilerplate code to quickly get started with development. The CLI also generates model, data sources, controller and repositories with boilerplate code based on user input. All this is generated based on the standard convention which means that it incorporates the API design best practices. (Developer, 2019)

*Figure 7 LoopBack Logo (Developer, 2019)*

Loopback is implemented in Typescript which is an open source superset of JavaScript. With typescript, the source code becomes easier to read and understand. It also provides static type checking which makes the development much efficient and productive. (Create REST APIs in minutes with LoopBack 4, 2018)

Loopback datasets provide data connectors for all the industry standard relational, NoSQL and Graph databases. Juglar is used as ORM/ODM with various drivers to deliver independence of the source code from the underlying database. The translation of them to the specific version and type of the database is managed by the loopback. This way you can change the underlying database system without modifying the source code. Loopback uses a code first approach for generating a database schema from the model classes in the application project using database migrations. (IBM, 2019) (Developer, 2019)

Loopback also provides a built-in API explorer and documentation generator. It allows you to visualize and interact with API resources without logic implementation. It makes both backend consumption and client-side implantation simpler and easy. It is fully customizable and environment independent which means it can run both locally and on the web. (IBM, 2019)

Loopback also has given plugins for JSON web token (JWT) authentication to secure the API endpoints. It has components such as sequence and interceptor which makes it easy to implement token-based API authentication.

### 1.1.5 Android SDK

The Android software development kit (SDK) (Wikipedia Community, 2018) (TutorialsPoint Community, 2016), powered by Google, is set of opensource tools used to develop apps that can run on the Android operating system. It consists of program libraries, emulator, debugger, profiler sample code, application programmable interface and support documentation for developing, debugging and profiling applications for Android. The android apps are developed natively in both Java and Kotlin. Kotlin is a modern statically typed language that executed in the Java Virtual Machine environment and offers complete interoperability with Java. (Meier, 2012)

Although with Android SDK command line tools, an app can be build using the command line, the Google provided a complete integrated development environment (IDE), Android Studio, for developing applications in a graphical interface environment. (Meier, 2012)

*Figure 8 Android SDK Logo (Meier, 2012)*

The Android SDK contains platform tools which are separately downloaded and contains command line tools including Android Debug Bridge (ADB). ADB is a command line tool to run commands on the android device connected to the system. The tool helps debug, deploy and fetch information about the app installed on the device. (TutorialsPoint Community, 2016) (Meier, 2012)

The Android Virtual Device (AVD) manager is used to create, manage and run android virtual devices also known as emulators. Emulators can be created with any android platform and can have configurable device hardware setting. The emulators are hardware neutral which means that it offers a better independent testing environment instead of a single test environment on a real device. (Meier, 2012)

The Android SDK ecosystem also supports and actively maintains backward compatibility on older devices. This helps the developer to target their applications on the legacy and old devices. This support is achieved using multiple support libraries that the SDK provides. (Academia EDU Community, 2008) (Meier, 2012)

SDK manager is an important tool of the Android SDK as it provides us the platform, resources and other useful components to build Android apps. It also manages SDK platforms and packages by downloading updates. (Meier, 2012)

### 1.1.6 Firebase

Firebase (Firebase Community, 2019) (Cheng, 2018) (Stonehem, 2016) (Moroney, 2017)is a web and mobile application development platform combined with tools and services that helps the development of the app, grow the user base and monetize the app. The following are a few of the services offered by Firebase SDK:

- Real-time Database: it a JSON based cloud-hosted real-time NoSQL database. It offers real-time syncing across multiple devices that helps user collaborate with each other.
- Authentication: it provides a state-of-art and easy to integrate authentication mechanism for the app. The authentication can be setup using a variety of mechanisms including email, google and twitter providing smooth sign-in and on-boarding experience.

- Cloud Messaging: it provides an efficient and robust push messaging service to your app. It delivers targeted messaging service to predefined segments, and groups of users based on demographics and behaviour.
- Remote Config: it is used to publish configurational updates to the devices. It is basically a key-value pair database where you can place your app configurations i.e. colour scheme, latest version and etc. it helps you to change the behaviour of the app without publishing the new version to the store.
- Test Lab: it is a service that offers multiple test mobile phones to test your app against real devices.it supports three modes of testing; instrumental test, robo test, and game loop test.
- Crashlytics: it is real-time, robust and lightweight crash reporting service that helps the developer discover, analyse and fix bugs and stability issues in the app. It also keeps track of the severity, occurrence, and priority of the crashes and highlights that to the developer accordingly.
- Performance Monitor: It helps your profile your application and provide insight into performance issues and characteristics. It also guides and reports the developer how and where can the performance of the app can. Be improved.

### 1.1.7 Braintree Payments

Braintree payment (Braintree, 2019) (Braintree Medium Community, 2019) is a global payment partner powered by PayPal. It is used to accept payment from the user via mobile and web interface. It is a full-stack payment platform and offers integration to other third-party payment gateways.

*Figure 9 Braintree Payment Logo (Braintree, 2019)*

The admin control panel offers a through a view of the transaction details and customer profiles. It also provides a wide range of checks and steps to avoid and prevent fraudulent transactions. There is also safeguard to check both the validity and status of the credit card before saving it in the Braintree customer account.

Braintree provides numerous integrations and payment methods including apple pay, visa checkout, ACH direct debit, master pass, PayPal, google pay etc. it offers both the customers and organizations to conduct and expand their business.

Braintree provides a sandbox environment and a control panel to test and debug our payment flow before deploying it out live. The control panel has test credit cards and accounts for all the possible use cases so that the user can test their app thoroughly in advance. It offers test transactions, refund and complete escrow process to test any possible purchase flow. (Websites 'N' More, 2018)

To integrate Braintree into the app, it provides both client-side and server-side SDKs. The client-side SDK is available for web, iOS and android. The server-side SDK is available for all the popular service-side platforms including Dot Net and Node.js. All the SDKs are opensource and publicly available on GitHub. The client-side SDKs provide both the building UI, which is called checkout UIs, and interfaces for creating custom UI for payment processing. (Abrosimova, 2015)

## 1.2 Design Decisions

In the implementation phase of the solution, I faced a lot of decision regarding the architecture and selecting between technologies that will contribute toward scalability and security of the system. Another important aspect was the overall usability of the system. The following section entails the important decision regarding the design and architecture of the proposed solution. In the next paragraph, I will provide the different phases of deciding upon a specific implementation or technology and the rationale behind it.

### 1.1.8 Blockchain vs Traditional Database

Traditional databases have been ruling the commercial data industry for four decades now. A traditional database is typically based on client-server and centralized architecture. All the data is stored and modified on a single monolithic centralized server by the database client. That forms the base for single point failure and data breaches. The control and authority of managing the database are also centralized is limited to a single participant. That is responsible to authenticate and provision a client's credentials before giving access to the system. This authority also monitors and control the administration of that database, which is centralized, the data can be altered if the authority is compromised. This means that any user with sufficient permissions can freely alter and append the data on the centralized server. (Silva, Almeida, & Queiroz, 2016)

Blockchain is a distributed ledger technology that is based on a distributed architecture and forms a network of several decentralised nodes. The network is a peer-to-peer network where each node is equally privileged in their control and access. Each and every node is capable of accepting new nodes to the network, verify their validity, accept and validate new transactions on the network. The heart of the blockchain is a consensus protocol that is responsible for reaching a consensus among the majority of that nodes while committing new transactions. That ensures the security and authenticity of the data. The blockchain works on a decentralized network scheme that means that there is no single point of failure or authority. It is also completely data immutable and new data is only inserted into the chain once the network reaches the consensus. (Zyskind, Nathan, & Pentland, 2015)

In a traditional database, there is always a trust issue among multiple parties and participant. Database Administrator (DBA) is a person or entity that manages and administers the database. That entity can alter the database for its own benefits and that too without a log or trace. Another possibility is that another party can influence the DBA into altering the data from the database for their personal gains. This is easily possible since it is a centralized architecture. In this system, different parties cannot form trust on the data. The major benefit of blockchain is decentralization which eliminates the central authority and control. It forms the basis of the trust where multiple parties can share information without a central authority. On a blockchain, each participant has a secure copy of all the transactions and a log that contains all the gradual changes being made to the ledger from each user. This transparency allows THE user to see all the data and its origin. The real power of blockchain comes in when there is a data inconsistency. The nodes in the network compare that inconsistency with the copy of their local data and immediately identifies and removes that data transaction. When the system is capable of self-data inconsistency correction and identification, it allows multiple parties to trust on it. The network is also able to track and indicate the user who tried to perform data alteration. The blockchain establishes data trust among multiple parties to function an perform their business. (Bauerle, 2018)

When multiple parties that are operating together are able to form the trust in the data that is being shared. It allows business enrichment and creates opportunities for more parties and participants to join the network. The data is secured by strong cryptography and security principles and is immutable. (Schlapkohl, 2019)

Another major difference between a traditional database and a blockchain system is log management. A traditional database only keeps the current data snapshot or in simpler words, it is an up to date data at a particular point in time. In contrast, a blockchain system has both the current snapshot of data and also the logs of all the data that has been inserted previously. In other words, a blockchain system maintains a historical record of all the transaction. This becomes very handful when tracing a record or an attempt to develop a data inconsistency. In a traditional database, this is not possible unless an external log tool is being integrated. (Rubikon Blockchain Corp, 2018)

The blockchain system is designed in a way that it appends new data to its network without affecting the previously added data. It allows us to see the current data as a formation of all the previously added blocks. As per the structure of the system, all the previously stored data cannot be altered or lost (Ray, 2017).

For our project, blockchain is an ideal choice. The parking logs will be generated at high volumes and there must be a robust and highly scalable system to cater to this data. Additionally, all the costing will be estimated on the check-in and check-outs of the parking logs. So, we require a system that can withhold adversaries and data inconsistencies. The system must also be decentralized to ensure trust on the data. Data will also be immutable so any participant cannot alter the past logs. This is essential for security purposes as well as it allows us to track car details and parking surveillance. One of the setbacks for using blockchain is performance. As compared to traditional databases, blockchain is relatively slow due to its distributed nature and decentralized control. this can be tweaked and adjusted to our needs. Our parking system will be based on a private ledger which means that strong encryption and a huge consortium of nodes will not be required. This is a trade-off and will significantly improve the performance of the blockchain transaction speed. Another query regarding the blockchain is the separation of heavy and light nodes. A heavy node has a complete snapshot of ledger data and takes part in the consensus. A light node is simply a node that can access and submit a transaction on the network. In our case, all the end users will be light nodes, as they will access their parking logs and payment transactions. All the parking terminals will be heavy nodes, they will keep the copy of the ledger data and will be connected to the network services.

### 1.1.9 Ubiquitous Computing

Ubiquitous computing (also known as pervasive computing) is a modern computer science phenomenon where computer sensors such as personal accessories connect with computing devices in our vicinity to provide an experience of ongoing connectivity and computing. It develops a network of interconnected sensor devices that makes the daily life activities of end-user fast, efficient and cost-effective. (Archetekt, 2017) (Zahrani, 2010)

It is brought into integration for its convenience and ease of use. It forms a human-machine interaction. In contrast to personal computing which focuses on activities of the participants, ubiquitous computing sensor devices focus to improve and facilitate our daily lives. That is why many businesses are following the concept of ubiquitous computing and following its concept. It is also implemented with AI and Machine Learning to leverage the potential of its modern approach. Its integration with other disciplines helps to establish a system which allows universal access and accessibility of computing functions. (Archetekt, 2017)

One of the issues with Ubiquitous computing is its creation of ethical issues regarding user consent and privacy. Computer systems in our daily lives are used consensually which is very different from pervasive computing that happens in our daily lives as it engages an individual without its consent. (Reference Community, 2019)

Ubiquitous computing is usually comprised of embedded sensor technologies that are usually implanted and out-of-sight. That makes the system gain more capabilities without visual clutter. It also promotes socialization in the technological community. Different devices are able to communicate and share their data to build better and usable computing. (MARIA LORENA LEHMAN, 2019)

Ubiquitous computing helps develop a smart environment that supports the users to make better and smart choices in their everyday daily lives. They are also self-sustainable to make a smart decision on behalf of users and leads to a proactive architecture. This also works as an architecture to develop into a nervous system for processing massive sensor data. The network gradually adapts from crunching data to analysing it to yield useful insights for better decision making. This self-learning system doesn't need constant inputs or parameter tweaking to function appropriately. (MARIA LORENA LEHMAN, 2019)

Ubiquitous Computing provides ease of use and persuasive mode of operation for our project. We will use Raspberry Pie as an IoT device to enable persuasive computing. The device will be placed on the vehicles and will be used for auto check-in and checkout at parking stations. The car will auto check-in when the car will reach near the parking station. The IoT device will connect with the parking lot terminal and log user details on the blockchain ledger. The logs will contain user data, vehicle information, and timestamp. When the user will leave the parking lot, the parking terminal will auto check-out the car. There will a computing device at the parking station to which all the IoT enabled vehicles will connect. This persuasive computing is not only safe as there is no personal information exchange and also time-saving. The user does not need to find any check-in counter or parking ticket at the parking station. That will not only save the time of the user but also this automated process will save the system from anomalies. From the end user perspective, the user will be able to configure the IoT device using their smartphone app. They can set up and connect their vehicle information to that device. After that, the device will log the check-in vehicle information to the blockchain ledger at the parking lot.

The computer in the parking lot will have a strong Wi-Fi field that will cover the whole parking areas. When an IoT enabled vehicle will enter into that field, the computer terminal will scan the IoT device and log its information in the ledger. The wi-fi terminal will continuously scan the field to make sure the vehicle is in the parking vicinity. When the scan can't find the vehicle, the parking terminal will log the vehicle out after a specified interval.

### 1.1.1 Credit Card vs Cryptocurrency

The word cryptocurrency was coined after the invention of bitcoin by Satoshi Nakamoto in his whitepaper. It is an electronic cash system for peer-to-peer transactions. It is a similar system of payments as of wire transfer where the money is transferred from one entity to another entity without the need of any financial intermediary. Cryptocurrency cards are used to utilise a huge variety of cryptocurrencies. It serves as a payment card for paying bills and withdraw cash as it allows the exchange of cryptocurrency into fiat currencies instantly (Broverman, 2018). A cryptocurrency debit card can be either virtual, physical or in the form of a prepaid coupon. It works exactly like a tradition debit or credit card except it doesn’t have a bank account or credit line behind it. It is connected with cardholders’ digital wallet where their cryptocurrencies are stored. It also provides a new kind of financial freedom as you can exchange out a variety of cryptocurrencies (Ethereum, Bitcoin, Litecoin and etc) as a source of fiat currencies (EUR, USD, PKR and etc). These cards also work on all Visa and Mastercard ATMs. Both allow withdrawing currency as fiat. (BLYSTONE, 2019)

Credit Card transactions authorize the seller to transact a payment from the buyers account into their account via financial intermediaries in the process. A typical card transaction comprises of four participants: the acquirer (an institution that delivers payment to the merchant), the issuer (the buyers/card holders bank), the merchant and the individual cardholder. (Broverman, 2018)

Cryptocurrency is decentralized and allows anonymous transactions. All the transactions are peer-to-peer and without any financial intermediator. That makes the transactions economical since there is no third-party involvement. The transactions also lack identity theft as it happens across borders. At the same time, cryptocurrency is extremely difficult to spend in the corporate world. Cryptocurrency transactions are also irreversible and can only be refunded via the receiving party. In contrast to credit card transactions, these are also anonymous, which means that no personal information, address or phone is attached with the transaction. This is a major security risk and can lead to money frauds. There are third-party players that mitigate this risk by acting as an intermediary but this costly and does not stop participants from dealing directly. The technology is still in its infancy and requires improved security measures for safe adoption. (Broverman, 2018) (Boukhalfa, 2017)

Another issue with cryptocurrencies is their lack of scalability. The number of multiple variants of cryptocurrencies is increasing rapidly but at the same time, the number of transactions collectively is less by what is transacted by VISA each day. The speed of transaction is also relatively slow as compared to financial master players i.e. VISA and Mastercard. To cater to these problems, an evolution for massive scaling of infrastructure is required. As of now, the financial institution has not shared any interest in this area. (Boukhalfa, 2017)

Cryptocurrencies also lack the stability in its values and suffer from price volatility. The whole ecosystem is recognized as a bubble that can burst anytime. This is also one of the major reasons for not adopting this technology for our project. For example, a customer purchased a parking ticket for 2 euro and its equivalent value in the cryptocurrency is 2 bitcoins. The next day if the value of the bitcoin changes, the customer might end up paying 5 bitcoins for the same ticket. This uncertainty loses customer confidence in using this payment system. (GROSSBERG, 2018)

# Chapter 4

# Solution

The whole system is a proof-of-concept (PoC) and comprised of several subsystems that work and connect with each other. It utilizes multiple technologies and frameworks to build the use case discussed in the previous sections. In this chapter, we shall understand the architecture of the system and its working. We shall also discuss the working and mechanism of each system in detail below.

## 1.3 Architecture

The system is composed of three subsystems that connect with one another using different communication protocols. The Web API Server is the main communication hub that connects with both Hyperledger Fabric Server (Blockchain) and Database Server (MySQL). The end-user communicates with the servers via android smartphone app and IoT device.

The WebAPI server is a NodeJS application which is responsible to connect with both blockchain and MySQL database. It has the set of APIs to store, retrieve and update user data on both the servers. Android App connects to the API server via HTTP/HTTPS protocol. The WebAPI server connects to MySQL and Blockchain server via TCP protocol. It also connects to Braintree Server for payment processing and Stripe Server for dispatching emails. It also leverages Firebase Services to dispatch push notifications to smartphone users. The WebAPI server is subscribed to blockchain server and actively listens to all the events emitted by it.

The blockchain server is responsible for storing parking logs and dispatching events on successful check-in and check-out. It has exposed WebAPIs to allows IoT devices to check-in and check-out the vehicle directly. The database server maintains a MySQL database instance and processes the WebAPI requests from NodeJS

Server. In the figure below, we can see the architecture diagram and all the components connected with each other.

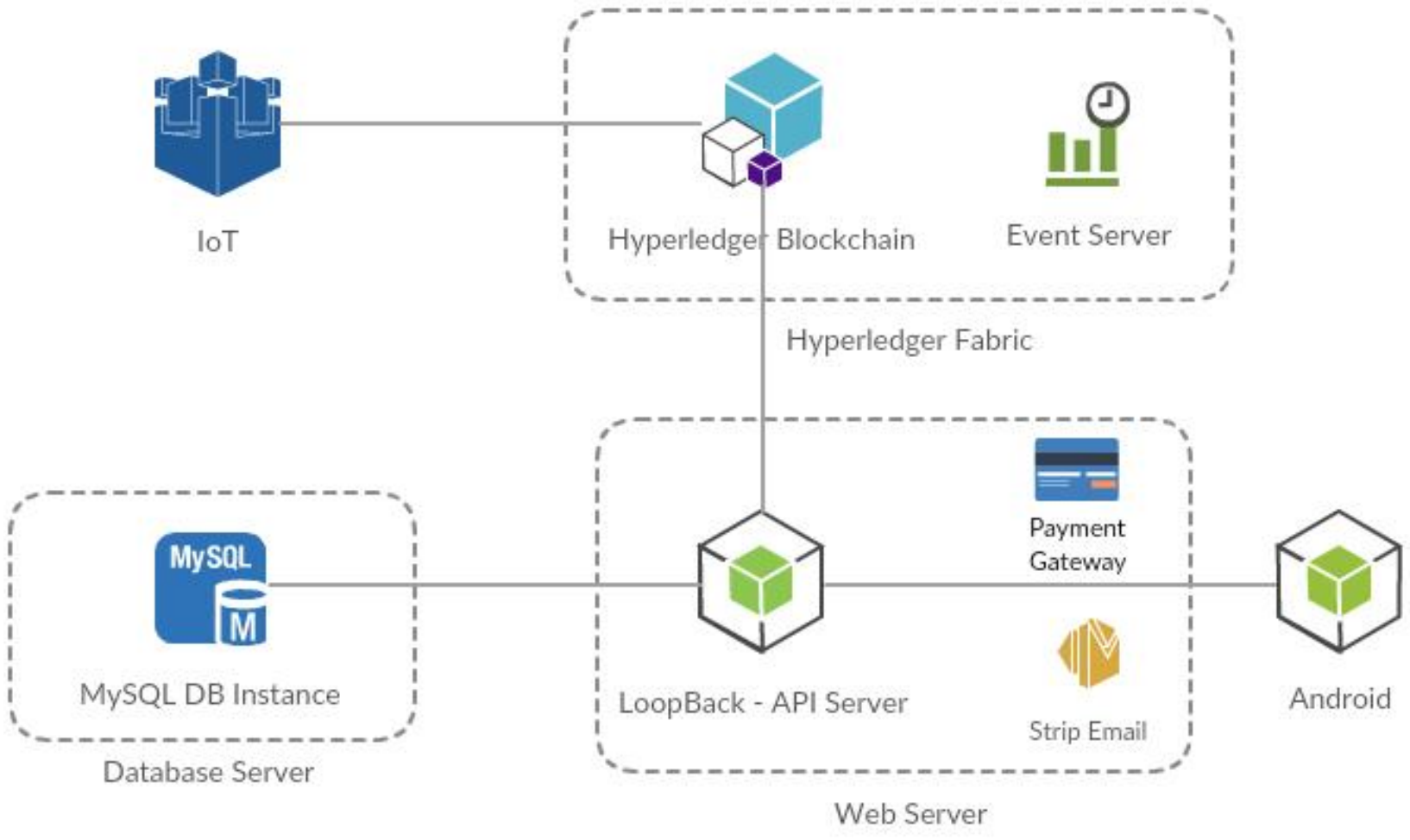

*Figure 10 Project Architecture*

In the upcoming sections, we will describe the working and components of all the technologies in detail.

## 1.4 Android App: Smart Parking

The android app has been developed for the end user to smartly obtain the parking space and process the payment. The app has been built using the latest Android SDK, which is API 28 – Pie (Developer - Android, 2019). The app utilizes AndroidX framework which is new support library framework introduced by android. The app is backward compatible until SDK 19 – Kitkat (Android SDK Offline, 2016). That makes the app available for around 98% of the whole operating devices. The app is built using material design which is lightweight and vector-based design scheme. The app is built using modern design principles and usability law. The app footprint size is only 3 MB and it requires no permission from the user.

In the section below, we shall discuss the working and functions of different modules of the app.

### 1.1.2 Login and Registration

In this module, the user is able to register himself on the app. After the registration, the user is asked to verify his account by submitting the pin code from his registered email. Users with unverified accounts will not be able to use the app.

After successful registration of the app, the user can log in to the app. For the first time users, the app will ask to setup driver information which will include driver's license information and registered name. In the next step, the app will ask the user to submit its credit card information to enable one-click quick payment of the parking ticket bills. Both the steps are optional for now and can be completed later. But it is crucial for the user to complete both of these steps to use the functions of the app.

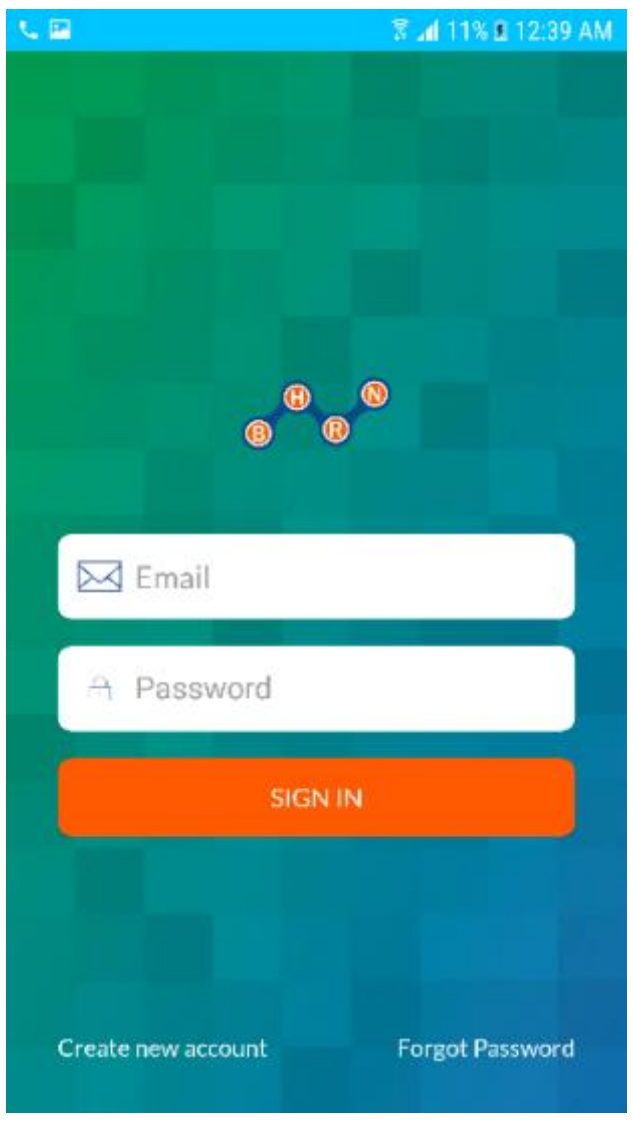

*Figure 11 Login Screen*

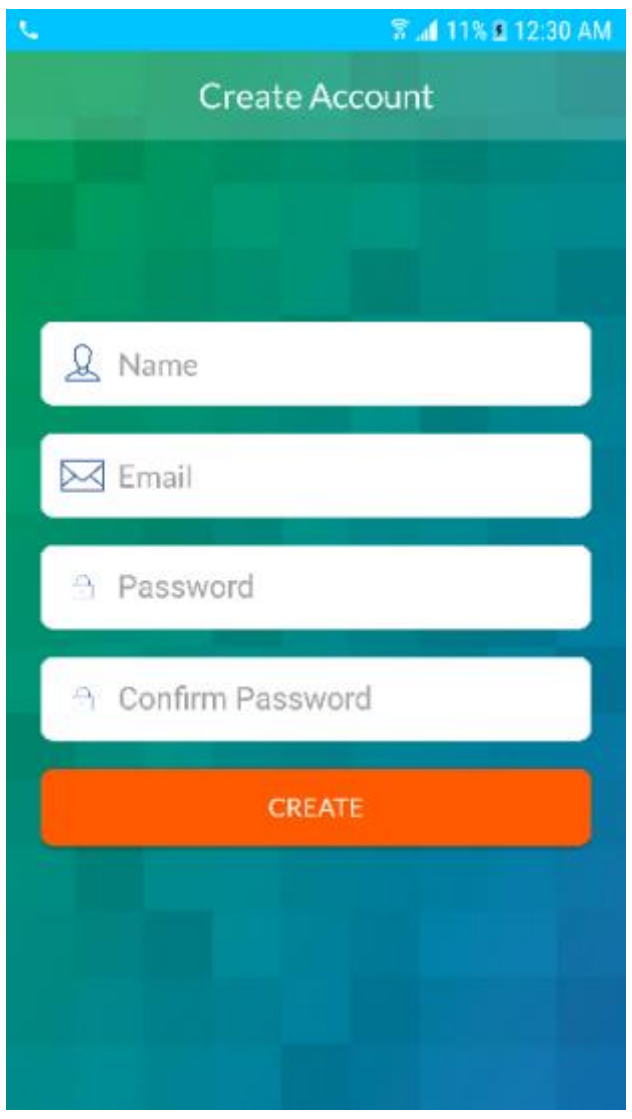

*Figure 12 Registration Screen*

Both login and register screen have field validation and checks for developing a strong password. After a successful login, the app fetches all the metadata for the user and also registers the phone at the firebase server to receive the push notification.

### 1.1.3 Vehicle Registration

In this module, the user is able to add, update and delete vehicles on his profile. When the user adds the vehicle, a unique code is generated that is configured on the IoT device. The IoT device uses that code for check-ins and check-outs. The android app uses this code from the vehicle to fetch parking logs and transactions. The vehicle adds dialog prompts the user to enter vehicle metadata such as model, make and plate number. There are all the essential validation checks to ensure correct information is being entered. The vehicle listing shows all the vehicle with swipe feature to enable deletion and update. Only one vehicle can be active at a time, which is indicated by active tag on the vehicle listing.

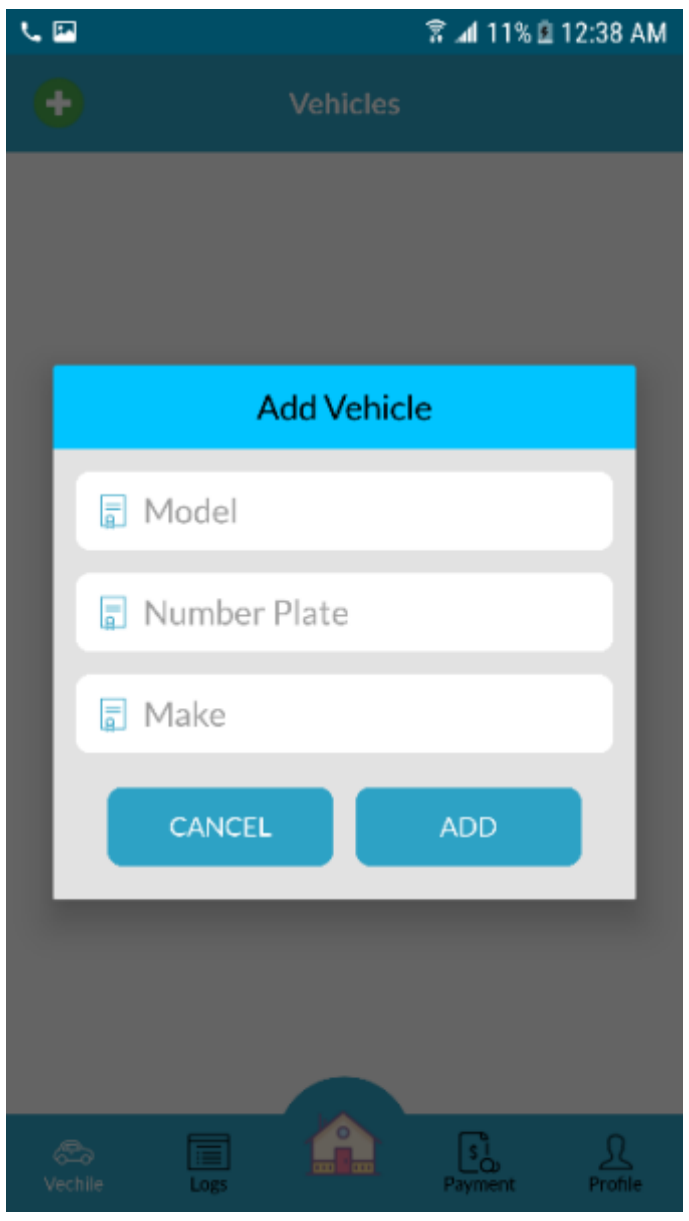

*Figure 13 Vehicle Add Dialog*

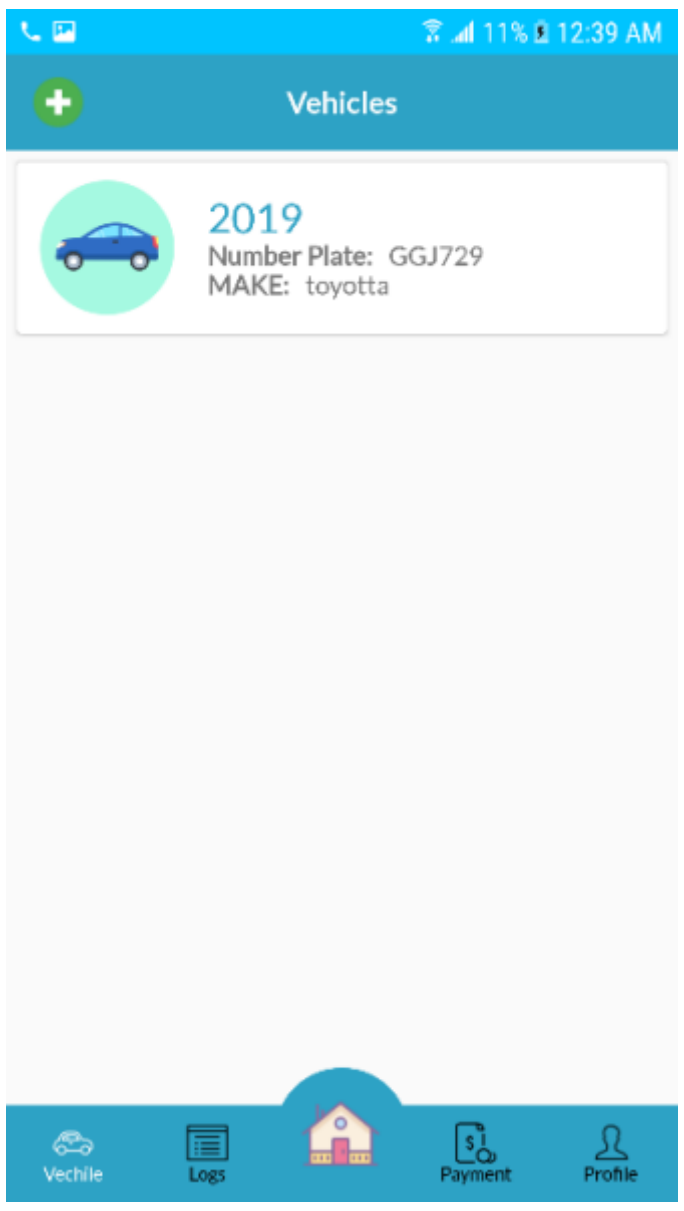

*Figure 14 Vehicle Listing*

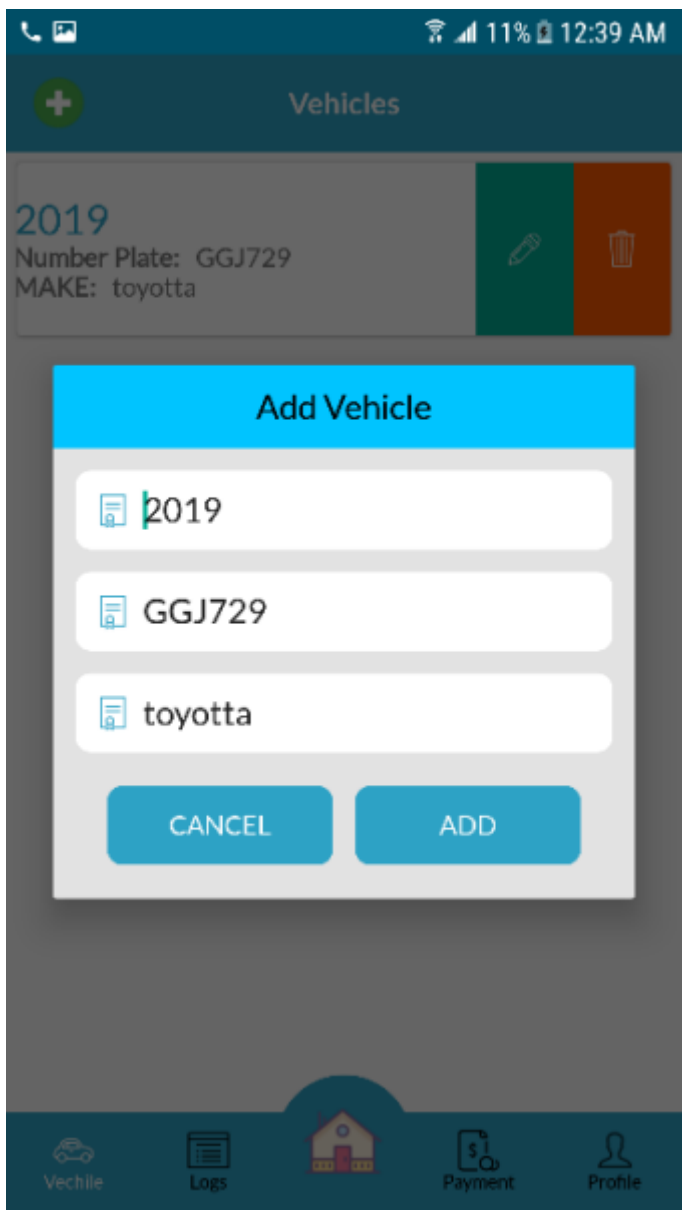

*Figure 16 Vehicle Update Dialog*

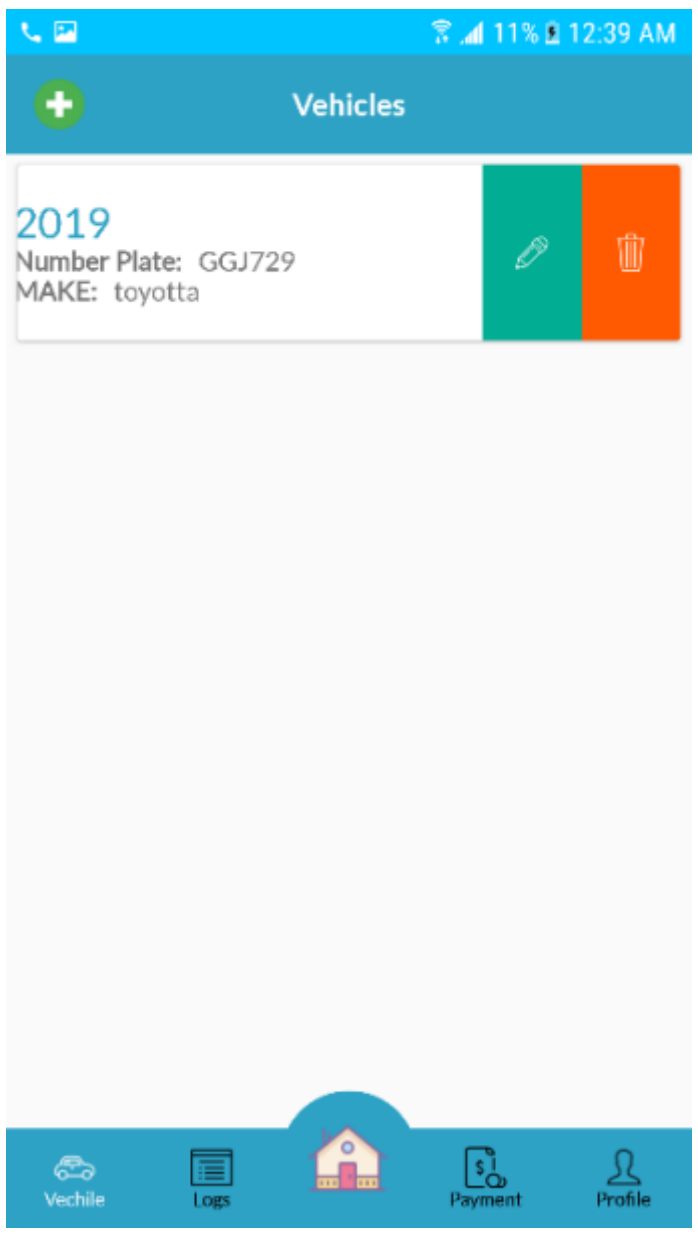

*Figure 15 Vehicle List Update*

### 1.1.4 User Profile Setup

In this module, users are able to set up their profile. They can also update their existing data including payment and driver information. The user profile screen shows user data. User can also log out from the app from this screen by clicking the logout button on the top left. It also has two more sections; driver and payment. In the driver section, the user adds their license information and in the payment section, user can register their credit card. The android app uses Braintree client-side SDK for credit card authorization on our server. This SDK also checks for fraud and invalid card and doesn't approve the card unless it is able to perform checking transaction.

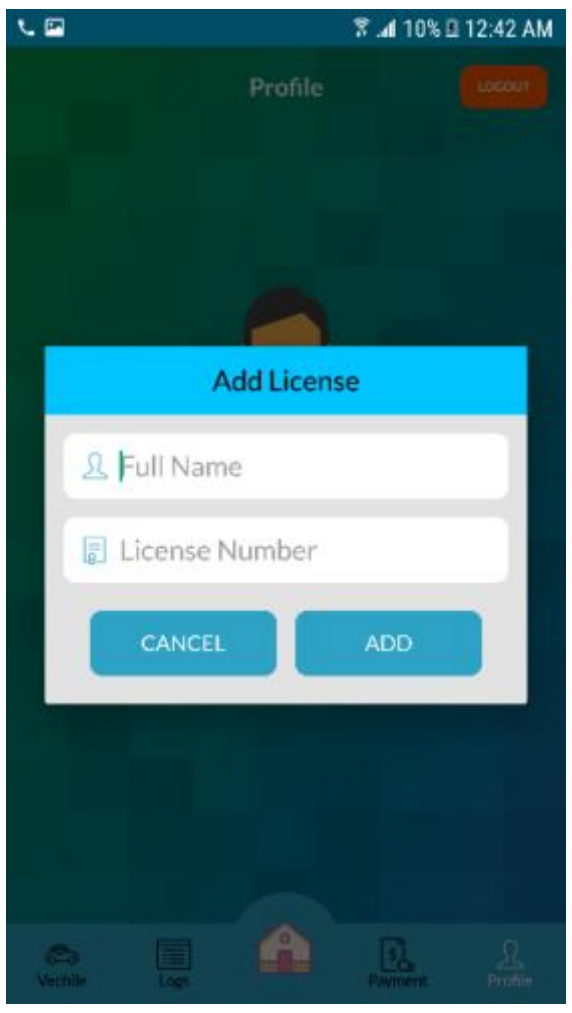

*Figure 17 Add License Screen*

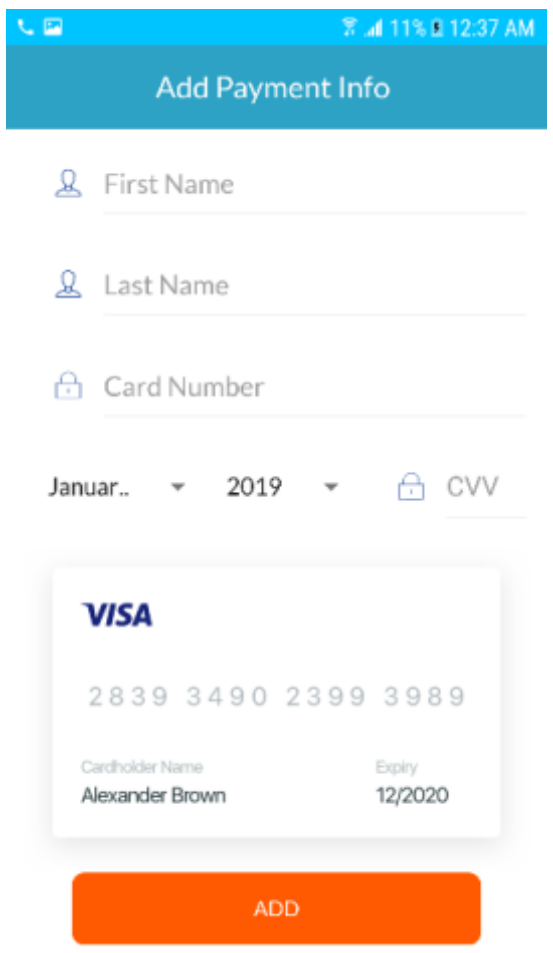

*Figure 18 Add Credit Card Screen*

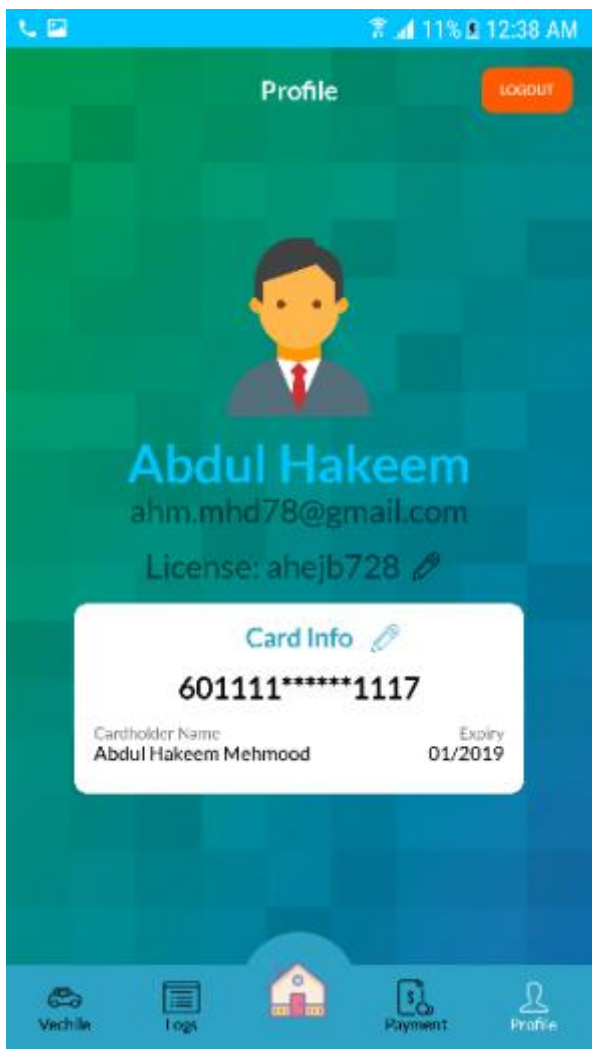

*Figure 19 User Profile Screen*

### 1.1.5 Parking Logs and Dashboard

In this module of the app, the user can check and verify all the check-ins and check-outs logs of their vehicle parking. Once the vehicle has checked out from the parking space, users are also able to see the total duration of their parking time and the associated cost. In the figures below, we can see two listings; one for parking logs and one for the transactions. User can also see the status of the ticket which can be either paid or unpaid. User can pay an individual ticket or all the unpaid tickets. The user also gets a push notification from the Firebase server for all check-ins and check-outs.

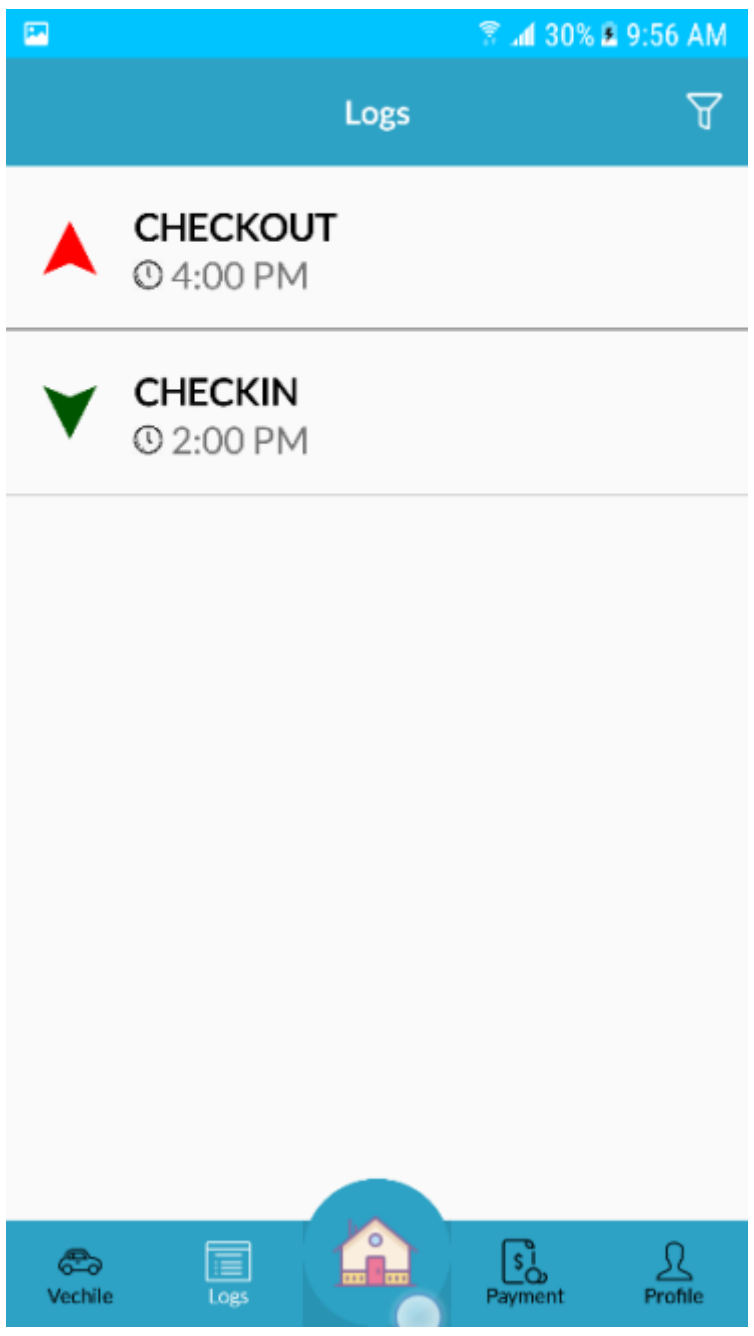

*Figure 20 Parking Logs*

### 1.1.6 Security

The app implements all the security guidelines provided by the android. The code is obfuscated and can't be reversed engineered. This protects the source code and all the API keys used in the project from unauthorized access. The app is enforced to use an SSL connection when available. All the permission in the app is signature based. The app manifest also does not contain any unused permission. The app uses shared preference excessively to store user data. The app opens share preference in private mode to avoid data leaks. The app uses all the up-to-date third-party libraries and dependencies. (Android, 2019)

### 1.1.7 Build and Install

Open the project in Android Studio and wait for the project to set up. Once it is launched, click on the Run > Run in the menu bar to build and launch the app. If your device is connected to the PC, it will install it on the device. Otherwise, it will open a dialog for "Select Deployment Target" to select the device. It gives an option for both physical devices or launching an emulator. You can monitor the whole build process by clicking View > Tool Windows > Build. It shows detailed building steps and output all the logs. You can profile the attributes of the app by clicking View > Tool Windows > Profile. It shows detailed statistics and data from memory, computing and rendering, and output all the logs. Based on the version of your Gradle, you can take advantage of the latest features offered by the Android Studio such as Instant Run, Build Variants and etc. (Android, 2019).

## 1.5 WebAPI Server

The webAPI server is a bridge between end users using android application and blockchain system that maintains the ledger of parking and transaction logs. It is built using IBM LoopBack 4 which is an API building framework on the top of Typescript and NodeJS. The Data Transfer Object (DTO) of the API is JSON, which is JavaScript object notation. The API is protected by JSON web token authorization which requires the client to authenticate first before accessing the API. It also supports role-based API access, but we have used only one role for this Proof-of-Concept. We have used MySQL database for storing metadata of the end user and blockchain for parking and payment logs. This system connects to both services. For the data maintained by the MySQL database, the LoopBack framework utilizes Juggler API for communicating with the database.

In the core, it is a node-based project and has all the configuration of a node project. The package.json file contains all the dependencies for both development and release environment. The major code base of the project has been programmed in Typescript except for few modules for which the typing files is not net yet developed. The password and other keys in the project have been encrypted using powerful hashing algorithms with a random salt.

In the sections below, we will explain the different modules of the API.

### 1.1.8 API Explorer

The Loopback is bundled with an API explorer which is very useful for testing and debugging the API. It is also used for developers to utilize the API as it provides extensive auto-populated documentation on input and output of the API. We can also place extra documentation such as error codes and their meanings, other parameters and notes. The developer can also test the API by executing it on the API explorer and analyse the output.

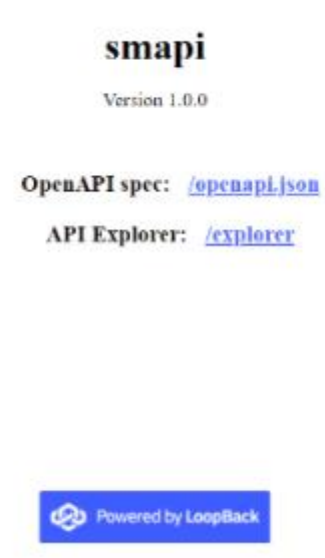

*Figure 21 API Explorer HomePage*

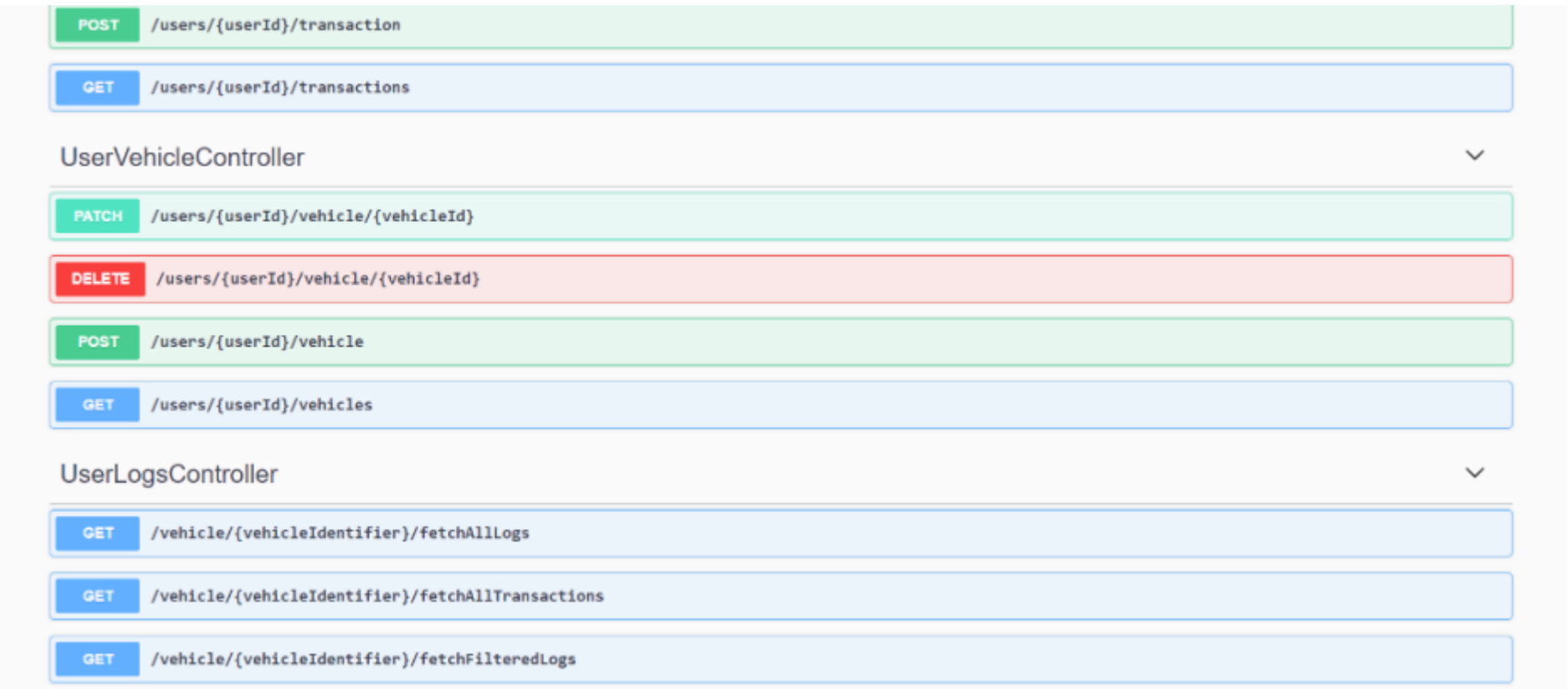

*Figure 22 API Explorer APIs*

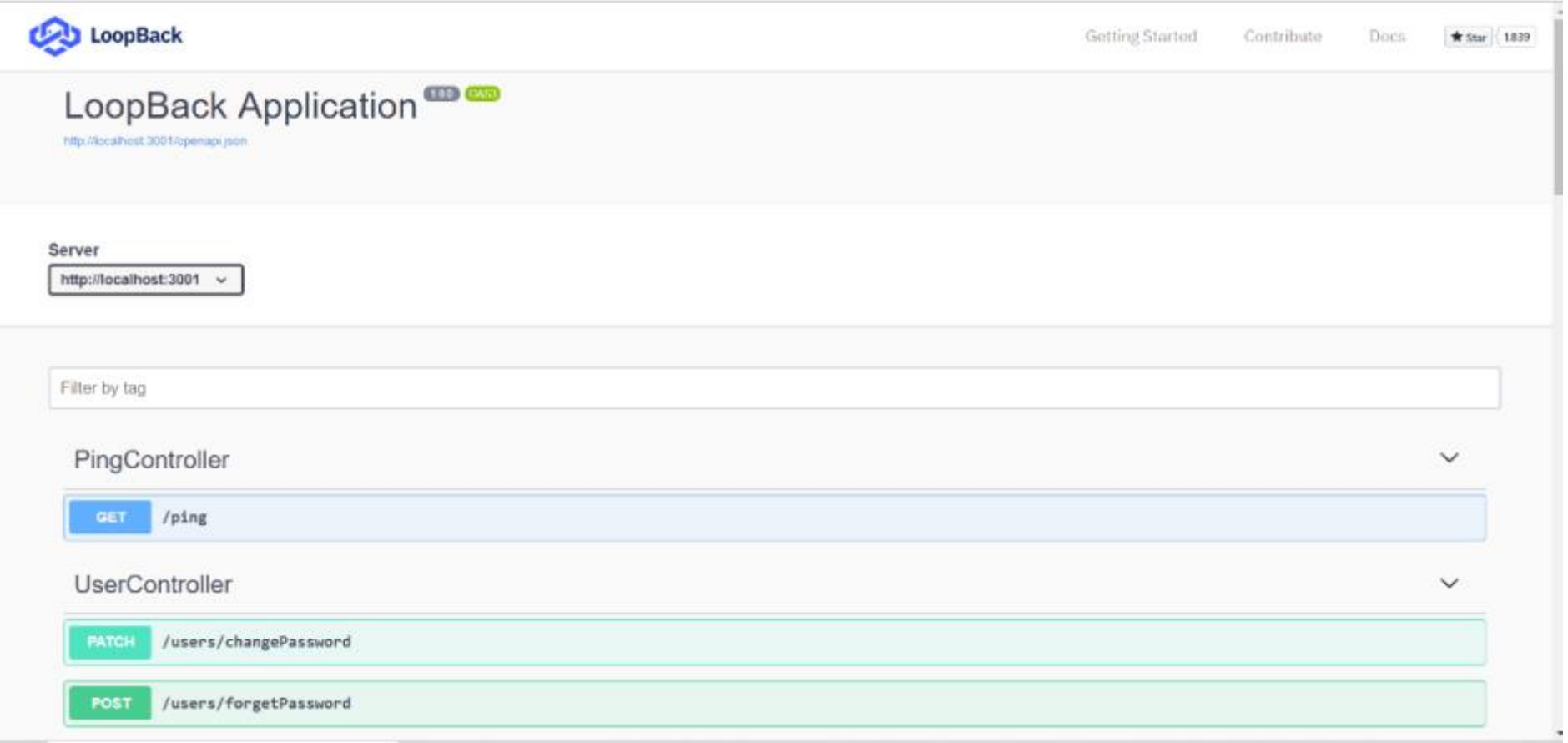

*Figure 23 API Explorer APIs*

### 1.1.9 Login and Registration

In this module, we have APIs for registration and authentication of the user. When the user successfully registers on our system, we use Stripe email service to send email to the registered email address with the account activation code. Same activation email flow is used in the forget password service. The user inputs the email address and receives a token on the email. After he verifies that token, he can change the password or assign a new password for the account.

After verification of the account, the user can login into the app. in the login response, the user receives all the data regarding vehicles, license information and parking logs of the current month. The password is protected by powerful hashing.

### 1.1.10 Parking Logs

This module contains APIs for fetching parking tickets data of active vehicles. When user check-in to the parking spot. The blockchain server registers user parking logs and emits an event after successful insertion of the record to the ledger. The API server actively listens to the event emitted by the blockchain server. When it receives the event, it fetches the data from the event data and sends push notification through the firebase server.

The input parameter for fetching parking logs and transactions is "Vehicle Identifier". After fetching the data from the blockchain server, the API server sorts the data and check for possible anomalies such as a missing check-out or check-in. The API server also checks for duplication of data. All of these are possibilities and could be caused by IoT interface. The API server opens up a TCP connection and communicates over sockets with the blockchain server. We have used an official NodeJS package for communicating with blockchain server.

### 1.1.11 Connecting to MySQL

The LoopBack connects to the MySQL database via DataSource. The models created in the project are translated into database entities. The process is called code first approach in which the database is constructed via model classes and relationships. Once the code is completed in the project, we can run migration commands provided by the LoopBack to create or update the database schema. The project uses the Juggler API to communicate with the database. The following figure shows the JSON file for the DataSource connection to the database.

```
{
  "name": "mysql",
  "connector": "mysql",
  "url": "",
  "host": "localhost",
  "port": 3306,
  "user": "root",
  "password": "",
  "database": "smi"
}
```

*Figure 24 JSON for the database data source*

### 1.1.12 Connecting to Blockchain

This module also remains connected to the blockchain server. The end user sends the request to the API server for fetching the parking logs. The API server sends that request to the blockchain server and fetches the logs from there. After that it process that data and send to the end user in JSON format. In LoopBack, you can connect to any third-party data service via Data Source Connectors (LoopBack Community, 2019).

The LoopBack utilizes a third-party NodeJS library to connect to the blockchain services. The typing of the library is not yet available so the code is written in plain JavaScript.

DataSource is a method for loopback to connect to different sources of data, which includes databases and APIs. On top of DataSource, LoopBack allows us to create a repository to provide access to the data. In figure 23, we have put the screenshot from the API project of the data source of the Blockchain RESTful API server. This is a JSON file with endpoints and metadata. The LoopBack service read the file and use it to provide access to the service.

Repositories are loopback's way of implementing specialized service interface for providing access and CURD operations over a database or service. In our project, we created several repositories for blockchain and database entities. It is built on the top of a DataSource to leverage data access. These services are then utilized by controller and service layers to build business logic.

```json
{
  "name": "restds",
  "connector": "rest",
  "baseURL": "http://localhost:3000/api/",
  "crud": false,
  "options": {
    "headers": {
      "accept": "application/json",
      "content-type": "application/json"
    }
  },
  "operations": [
    {
      "template": {
        "method": "GET",
        "url": "http://localhost:3000/api/queries/getAllParkingLog",
        "query": {
          "uid": "{uid}"
        }
      },
      "functions": {
        "getLogs": [
          "uid"
        ]
      }
    }
  ]
}
```

*Figure 25 DataSource JSON for Blockchain API*

### 1.1.13 Connecting to Braintree Payment Gateway

The API also handles all the payment related task by utilizing the Braintree NodeJS library (Braintree Community, 2019). It is available as an NPM package from the official website. The API handles credit card insertion and updating. The API also performs all the checks for a fraudulent card using the library. The payment for parking tickets also takes place using the Braintree library. In the figure below, we can see the code snippet from our project for adding a new credit card and associating it with the user. On error, the error message is returned to the API response.

```
var promise = new Promise<CardResponse>((resolve, reject) => {
  gateway.customer.create({
    firstName: "Charity",
    lastName: "Smith",
    paymentMethodNonce: req.token
  }, async function (err, result) {
    if (result.success) {

      card.autoFill(result.customer!);
      output.card = await this.userRepository.creditCard(userId).create(card);
      output.response = new BaseResponse(1, "Credit Card Added");
      resolve(output);
    } else {
      output.response = new BaseResponse(-1, err.message);
      reject(output)
    }
  });
});
return promise;
```

*Figure 26 credit card code*

### 1.1.14 Security

Security in a project is essential and is often overlooked. There are two security systems built in the project. One of them is a secure password hashing. A service has been developed to securely hash the plain text password and further enhancing the security of hashing by appending salt. Hashing is a special type of algorithm which takes a variable length plain text as input and transforms it in fixed length output. It is hard and infeasible to reverse the hash and fast to compute for any given text. That is why it is also a one-way function. We have used bcrypt.js (Wirtz, 2017) NPM package for generating hashing for the password. It is essential for passwords to be hashed as it protects it in the form that even if it is compromised, it makes no sense. At the same time is very easy to verify the password as computing hash is very fast. The special thing about bcrypt.js is that it allows repetitive hashing in the form of rounds. The higher the number of rounds, the higher the cost of data processing and computing. Even if someone recovers the hashed text, it will be very difficult to retrieve the plain text through brute force strategy.

```
export type HashPassword = (
  password: string,
  rounds: number,
) => Promise<string>;
// bind function to `services.bcryptjs.HashPassword`
export async function hashPassword(
  password: string,
  rounds: number,
): Promise<string> {
  const salt = await genSalt(rounds);
  return await hash(password, salt);
}

export interface PasswordHasher<T = string> {
  hashPassword(password: T): Promise<T>;
  comparePassword(providedPass: T, storedPass: T): Promise<boolean>;
}
```

*Figure 27 Hashing Protocol*

The other security protocol implemented in our project is API authentication. Our API allows access to a large amount of data that involves privacy concerns. If anyone is able to query the data of anyone else, they will know about their whereabouts and location traits. To resolve this privacy concern, we have implemented API authentication in our project. The user who wants to access the API, he must first authenticate using his login credentials. In the response, he receives an auth token. For any subsequent API calls, he will need to place that token in the header of the request. Only the request with a valid token will be processed. Using this token, the user will only be able to obtain his own data. This process ensures that the end user can access only his data and not anyone else's.

In our project, we have used LoopBack JWT authentication packages to implement API authentication. This authentication service is injected into the sequence module of LoopBack system.

The sequence module configures how a web API request shall be processed. In the figure below, it is a part of JWT strategy from our project. JWT stands for JSON web token, which is a popular API authentication strategy. We can see that the requests without a token are returned with a server error code, Code: 500. If the request has a token that starts with Bearer, the token is verified for its validity first, after that the API request is forwarded to the controller.

```
export class JWTStrategy implements AuthenticationStrategy {
  constructor(
    @inject(JWTAuthenticationBindings.SERVICE)
    public jwt_authentication_service: JWTAuthenticationService,
    @inject(JWTAuthenticationBindings.SECRET)
    public jwt_secret: string,
  ) {}
  async authenticate(request: Request): Promise<UserProfile | undefined> {
    let token = request.query.access_token || request.headers['authorization'];
    if (!token) throw new HttpErrors.Unauthorized('No access token found!');

    if (token.startsWith('Bearer ')) {
      token = token.slice(7, token.length);
    }

    try {
      const user = await this.jwt_authentication_service.decodeAccessToken(
        token,
```

*Figure 28 JWT Authentication Strategy*

### 1.1.15 Connecting to Firebase

The project connects to firebase service for dispatching firebase notifications to mobile users. The project uses the firebase NodeJS SDK for connecting to the firebase service. The API server must have internet access to obtain and utilize firebase services. In the figure below, we can see that we have registered our server for receiving events from the blockchain. On receiving events, the data from the event is extracted and passed to push notification method for generating a push notification. In the push notification method, the SDK is initialized by the JSON configuration file obtained from google firebase project. After that push notification payload is being created for dispatching.

```
    ...
    this.bizNetworkConnection.on('event', (evt) => {
        console.log("â€Šâ€”â€Šâ€”â€Šâ€”â€Šâ€”â€Šâ€”â€Š-event happendâ€Šâ€”â€Šâ€”â€Šâ€”â€Šâ€” -");
        console.log("uid: " + evt.uid);
        console.log("time: " + evt.time);
        console.log("type: " + evt.type);

    });

}


export function sendPush(to: string, content: string, subject: string) {

    var serviceAccount = require("../smart-parking-firebase-adminsdk.json");

    admin.initializeApp({
        credential: admin.credential.cert(serviceAccount),
        databaseURL: "https://smart-parking-c796f.firebaseio.com"
    });


    var registrationToken = to;

    var message = {
        data: {
            message: content,
            subject: subject
        },
        token: registrationToken
    };

    // Send a message to the device corresponding to the provided
    // registration token.
    admin.messaging().send(message)
        .then((response) => {
            // Response is a message ID string.
            console.log('Successfully sent message:', response);
        })
```

*Figure 29 Push Notification Code Snippet*

### 1.1.16 Start-up the Server

To launch the server, the first step is to install all the project dependencies listed in the project.json file. You can do this by executing the following command in the terminal of the root directory of the project:

*$ npm install .*

After that, you will need to execute the database migration. This is achieved by executing the following command.

*$ npm start migrate*

Next, we will build the project for generating the binaries. To build the project, execute the following command:

*$ npm run build*

Finally, launch the project on the localhost at port 3000 using the following below command. Note that both the IP address and port of the project is configurable in the app.js file at the root directory of the project.

```
$ npm run start
```

The project will be launched at 127.0.0.1:3000. To browse the API, go to browser and type: http://localhost:3000.

## 1.6 Hyperledger Composer

Hyperledger Composer provides a set of tools to build distributed Apps – DApps on Hyperledger fabric. It is an opensource framework to develop and accelerate blockchain applications and its integration with the existing enterprise technologies. The composer has modelling technologies to develop and deploy our business use cases on business networks. To setup composer, environment and runtime, you need the following set of dependencies. After that, you can use composer command line tools to create the composer project. We have used Visual Studio Code (Microsoft, 2019) for building, executing and debugging composer apps. A composer project consists of several components which combine to leverage distributed app capabilities. In the section below, we will discuss each of them with code snippets from our project.

### 1.1.17 Architectural Components

Hyperledger Composer works on the modelling paradigm where it allows to quickly model the business use cases using assets, participants and transactions. Assets in a business network can be either a tangible good or an intangible service. In our project, we have declared timesheet logs as intangible assets of the network. In the figure below, we can see the schema of the 'logs' asset.

Participants are users of the blockchain network that access or modifies the assets of the ledger by executing transactions. There could be more than one participant in a business network depending upon the business network requirements. For our project requirements, we have created one participant type, Driver.

Each driver in our system is associated with a driver profile on MySQL. A driver can create a parking timesheet log and modify its content by executing transactions on the business network.

```
enum Action {
  o CheckIn
  o CheckOut
}

enum RegionClass {
  o DB
  o LN
  o CK
}

asset Timesheet identified by id {
  o String id
  o String uid
  o DateTime time
  o Action actionType
  o RegionClass region
}
```

*Figure 30 Schema for Timesheet model*

### 1.1.18 Querying the Blockchain

Hyperledger composer provides a bespoke query language to draft queries in the blockchain. The queries are written in a file called queries.qry and is the part of the business network project. For our project, we had a few query requirements which are detailed below.

The first query was to obtain the timesheet logs from the ledger for a particular vehicle. In the figure below, we can see that the query takes a vehicle identifier as input and use it in the query command to filter results. This is an important query and the result is transported to the end-user mobile app via web API server.

```
query getAllParkingLog {
  description: "Get all parking logs"
  statement:
    SELECT org.avialdo.smartt.timesheet.Timesheet
      WHERE (uid == _$uid)
}
```

*Figure 31 Get Parking Logs Query*

Another important query was to fetch the last check-in from the ledger for a particular vehicle. This is required when a vehicle checks out and we want to calculate the total parking time and cost of parking. This is only possible if we have the last check-in log from which we can determine the time difference and hence the cost of parking. in the figure below, we can see the query code for fetching the last parking log.

```
query getLastCheckedInTransaction {
  description: "Get last checkedIn transaction"
  statement:
    SELECT org.avialdo.smartt.timesheet.Timesheet
      WHERE (uid == _$uid  AND actionType == 'CheckIn')
        LIMIT 1
}
```

*Figure 32 Get the Last Transaction*

### 1.1.19 Transactions

Transactions, also known as chain code, are a mean of executing data on the ledger. There is a different kind of transaction functions such as asset transaction, participant transaction etc. We have used only the asset transaction function in our project to insert the timesheet check-in and check-out logs. In the figure below, we have logic from our logic.js class that has an implementation of transactions.

```
async function checkIn(tx) {
    const factory = getFactory();
    const NS = 'org.avialdo.smartt.timesheet';

    const id = new Date().getUTCMilliseconds().toString();
    const timesheet = factory.newResource(NS, 'Timesheet', id);
    timesheet.region = tx.region;
    timesheet.uid = tx.uid;
    timesheet.time = new Date();
    timesheet.actionType='CheckIn';
    // save the order
    const registry = await getAssetRegistry(timesheet.getFullyQualifiedType());
    await registry.add(timesheet);

    // emit the event
    const checkEvent = factory.newEvent(NS, 'CheckInEvent');
    checkEvent.uid = timesheet.uid;
    checkEvent.time = timesheet.time;
    checkEvent.type = 'CheckIn';
    emit(checkEvent);
}
```

*Figure 33 Check-in Transaction Code*

When a transaction is executed by a peer in the network, it undergoes a process of approval and validation before it is committed to the ledger. Each network has an endorsement policy that states how the transaction is going to be approved. Specifically, it states that which peers of the network or at least how many peers from the network must validate a transaction before it is added to the ledger. When a peer sends a transaction request, it must fulfil the endorsement policy. Endorsement policy is controlled an Orderer, a node responsible for the whole transaction approval process. After the fulfilment of endorsement policy, the orderer notifies all the peers to add the transaction in their local ledger. The orderer itself adds the transaction in the world ledger or global ledger. If during the endorsement process, the transaction is rejected by any of the peers, it is not added to the ledger. In the endorsement policy, we can apply various validation rules to avoid certain threshold, duplication or checking any prerequisites.

### 1.1.20 Events

Hyperledger composer emits events that can be subscribed by third-party applications. It provides a means to share data and updated to other applications in the enterprise. The definition of the event is described in the model file and its implementation is covered in the logic.js file. It is up to the developer to publish the event in the transaction logic. For our project, we have created to events namely, check-in and check-out. The events as the name suggest are triggered when a vehicle is check-in or check-out. The events are registered in NodeJS Application that is our API project. In the figure below, we can see the event emitting code and business logic.

```
event CheckInEvent {
  o String uid
  o DateTime time
  o String type
}

event CheckOutEvent {
  o String uid
  o DateTime time
  o String type
}
```

*Figure 34 Event Definition*

### 1.1.21 Permission and Privilege

In Hyperledger Composer, we are provided with an Access Control Language (ACL) (Hyperledger Community, 2019) which allows us to declare access controls for participants over the business network elements. ACL allows us to define roles to users on the create, update and delete operation over elements in the business network. The rules use strict namespaces to identity business network resources and are written in an access control file (.acl).

For our project, we used the default permission rules defined by the hyperledger. We defined two types of participants, network and system administrator and granted all the privilege to their scope. In the figure below, you can find the definitions of our ACL file.

```
rule Default {
    description: "Allow all participants access to all resources"
    participant: "ANY"
    operation: ALL
    resource: "org.avialdo.smartt.**"
    action: ALLOW
}

rule SystemACL {
    description:  "System ACL to permit all access"
    participant: "org.hyperledger.composer.system.Participant"
    operation: ALL
    resource: "org.hyperledger.composer.system.**"
    action: ALLOW
}
```

*Figure 35 ACL File*

## 1.7 IoT – Raspberry Pie

In this module, we want to accomplish and develop a system that would utilize the IoT device and implement the paradigm of ubiquitous computing for auto check-in and check-out.

This module of our project is not implemented and is not in the scope of our project. What we have done to demonstrate our proof-of-concept is to develop a python program that will auto check-in and check-out randomly. Although, we have an idea for the working of the IoT network and the mechanism for auto check-in and check-out. it is inspired by a work of students from the MIT (Nadeem, 2017) using Wi-Fi probe. In the work, they used to detect the check-in and check-out of people in the MIT corridor using Wi-Fi probes. The main idea was to detect the wi-fi radio receiver of people's cell phone. The wi-fi receiver constantly sends wi-fi to connect packets for the router to receive and to connect. These wifi packets contain the MAC address of the device along with other useful data that helps to identify a device uniquely. The main router that keeps scanning the devices receives the wifi probe and opens it and marks the check-in after verifying the mac address from its lookup table.

We can use the same strategy and implement the check-in and check-out of our system. The parking terminal at the parking lot will have an IoT device with a wifi router. The range of the router will cover the whole area of the parking lot. When a vehicle will enter the parking lot, its IoT's wifi router will send the custom wifi-probe with the vehicle identifier. The parking terminal will mark the check-in of the vehicle. It will continuously scan the lot for the vehicles to ensure that they are part of the wifi field. If a vehicle does not respond for a particular time. The parking terminal will mark the check-out of that vehicle. This approach is still conceptual and requires implementation and testing for its feasibility and verification.

# Chapter 5

# Conclusion and Future Work

In this chapter, we summarize this study, draw a conclusion, and suggest further work.

As mentioned in the first and second chapter, the world traffic is getting congested gradually, and the population especially in the urban areas are facing serious parking problems. A lot of research and experimental work has been accomplished in the past to implement a smart and efficient parking system, but there has not been any concrete and major implementation on it. It is a grave issue and affects people daily, especially those who commute to work in the city centres and surrounding vicinity. It is routine for the commuter to find a parking space and it has been estimated that 30% of the urban traffic congestion is caused by drivers spotting for parking spaces. Also, the fraud committed by people who reuse the old parking tickets leads to huge financial loss to the government. There is no tracking of vehicle parking which otherwise could benefit agencies to track and avoid criminal activities. It is an important part of urban planning and requires a global and standard solution. If the major and metropolis cities fail to migrate towards smart parking, it will be a major issue for their sustainability and future extension.

In this thesis, we first introduced the problem and our project objectives. After that, we surveyed the literature on the previously developed smart parking solutions from an academic and technical perspective. we also reviewed why different proposal and solution could not be integrated and which system has the potential to cater to this problem. Then we proposed a novel solution for implementing smart parking using cutting edge technologies. We surveyed and explored all the technologies in the third chapter of this thesis. In the same chapter, we discussed and reasoned our design decision for selecting a certain technology over another. We also discussed the factors behind build this project on the principles of pervasive computing. These design decisions also summarized the advantages and disadvantages of different technologies and how will they integrate with one another. In the fourth chapter, we presented our solution, Smart Parking. It is developed using cutting edge and modern technologies. It will facilitate the drivers to spot, acquire and pay for the parking places. The ability of the system to avoid direct cash exchange or parking ticket purchase will provide ease to the drivers. The system works on the principles of pervasive computing and provides auto check-in and check-out. The parking logs and transactions are secured by the blockchain system. The user can control the system and their profile using the app on their smart phone. Another advantage of the system is easy and online payment method. User can pay for their parking tickets using their credit card from their smart phone app. This decreases their hassle of carrying cash and coins for purchasing parking tickets. Smart Parking will optimized the parking mechanism, save time, reduce traffic and pollution, and provide an enhanced user experience.

We combined blockchain and IoT to develop a proof of concept (PoC) of smart parking solution. Both the technologies are in their nascent form and require maturity and community support for large scale and enterprise implementation. During the implementation of the PoC, we faced a lot of technical problems using and integrating these technologies to make the system work. There were very less community support and most of the times we had to rely on trial and error strategy for coming up with a solution. One such issue that required significant efforts was to develop and integrate typing for the Braintree payment SDK. The Braintree payment SDK does not have official typing for its functions. Based on an open-source typing version (Shahine, 2019), we developed the typing for the SDK and implemented it in our code. Lacking strong community support is a big challenge and a limitation when working on cutting edge and nascent technologies. It has lacked the design pattern and best practices for enterprise implementation. Another limitation of the system is to enable the user to perform manual check-in. If the user can perform manual check-in, they can hold a parking spot and sell it for more money. The current system implementation does not have the functionality to stop this act.

As for future work, the most important task is to implement the IoT module. As not being part of the scope, it still lacks proper functioning and implementation. The implementation and testing of this feature will determine the accuracy and effectiveness of wi-fi probe methodology for similar problems as well. Another important aspect of this system is to transform this proof of concept to an enterprise-grade application. Currently, the blockchain implementation is not purely distributed, and different nodes are active on the single work station using Docker containers. We need to deploy and test the blockchain system on multiple machines.

The system performance will show different parameters when it will be deployed on multiple machines. The network latency and communication congestion among multiple nodes affect the performance and speed of the ledger transactions, which ultimately affects the whole system. Many people park vehicles simultaneously at different places in the city and the enormous number of check-ins and check-outs will produce high-velocity data. Therefore, it is required to test the performance of the blockchain system facing this high velocity of data from the IoT device. It is also required to determine the implantation cost and feasibility of implementing this system to provide smart parking.